\documentclass{article}
\usepackage{arxiv}

\usepackage[utf8]{inputenc} 
\usepackage[T1]{fontenc}    
\usepackage{hyperref}       
\usepackage{url}            
\usepackage{booktabs}       
\usepackage{amsfonts}       
\usepackage{nicefrac}       
\usepackage{microtype}      
\usepackage{lipsum}		
\usepackage{graphicx}
\usepackage{doi}
\usepackage{natbib}
\usepackage{amsmath}
\usepackage{verbatim}
\usepackage{colortbl}
\usepackage{xcolor}
\usepackage{placeins}
\usepackage{adjustbox}
\usepackage{bm}
\usepackage{multicol}
\usepackage{multirow}
\setcitestyle{numbers}
\setcitestyle{square}
\date{}

\newcommand{\modif}[1]{\textcolor{black}{#1}}  

\title{\fontsize{15}{13}\selectfont \modif{Assessing data-driven predictions of band gap and electrical conductivity for transparent conducting materials}}

\author{Federico Ottomano$^1$\thanks{Corresponding author: \texttt{federico.ottomano@liverpool.ac.uk}}  , John Y. Goulermas$^1$\thanks{Deceased in May 2022}  , Vladimir Gusev$^{1,2}$, Rahul Savani$^{1,3}$,  Michael W. Gaultois${}^2$, \\ \\ \textbf{Troy D. Manning${}^2$, Hai Lin$^2$, Teresa Partida Manzanera${}^2$, Emmeline G. Poole${}^2$, Matthew S. Dyer${}^2$, } \\ \\  \textbf{John B. Claridge${}^2$, Jon Alaria${}^2$, Luke M. Daniels${}^2$, Su Varma${}^4$, David Rimmer${}^4$, } \\ \\ \textbf{Kevin Sanderson${}^4$, Matthew J. Rosseinsky$^2$} \\ \\
	${}^1$Department of Computer Science \\  University of Liverpool, Ashton Street L69 3BX \\ Liverpool, UK\\ \\ 
	${}^2$ Materials Innovation Factory, Department of Chemistry \\ University of Liverpool, 51 Oxford Street L7 3NY \\ Liverpool, UK\\  \\
	${}^3$ The Alan Turing Institute \\ British Library, 96 Euston Rd. \\  London NW1, UK \\ \\ 
	${}^4$ Pilkington Technology Management Ltd. \\ NSG Group European Technical Centre \\ Hall Lane, Lathom \\ Ormskirk L40 5UF, UK
}

\renewcommand{\shorttitle}{\textit{arXiv} Template}

\hypersetup{
pdftitle={A template for the arxiv style},
pdfsubject={q-bio.NC, q-bio.QM},
pdfauthor={David S.~Hippocampus, Elias D.~Striatum},
pdfkeywords={Machine Learning, Materials Discovery, Materials Informatics},
}

\begin{document}
\maketitle
\begin{abstract}
Machine Learning (ML) has offered innovative perspectives for accelerating the discovery of new functional materials, leveraging the increasing availability of material databases. Despite the promising advances, data-driven methods face constraints imposed by the quantity and quality of available data. Moreover, ML is often employed in tandem with simulated datasets originating from density functional theory (DFT), and assessed through in-sample evaluation schemes. This scenario raises questions about the practical utility of ML in uncovering new and significant material classes for industrial applications. Here, we propose a data-driven framework aimed at accelerating the discovery of new \textit{transparent conducting materials} (TCMs), an important category of semiconductors with a wide range of applications. To mitigate the shortage of available data, we create and validate unique experimental databases, comprising several examples of existing TCMs. We assess state-of-the-art (SOTA) ML models for property prediction from the stoichiometry alone. We propose a bespoke evaluation scheme to provide empirical evidence on the ability of ML to uncover new, previously unseen materials of interest. We test our approach on a list of 55 compositions containing typical elements of known TCMs. Although our study indicates that ML tends to identify new TCMs compositionally similar to those in the training data, we empirically demonstrate that it can highlight material candidates that may have been previously overlooked, offering a systematic approach to identify materials that are likely to display TCMs characteristics.
\end{abstract}


\section{Introduction}
Data-driven approaches have proposed a valuable change of perspective in the discovery of new functional materials, assisting traditional methods based on experimental investigation and density functional theory (DFT) calculations \citep{from_dft_to_ml, data_driven_mat_design}. This has been made possible by the consistent growth of available material repositories (Materials Project \cite{matproj}, Materials Platform for Data Science \cite{mpds}, Open Quantum Materials Database \cite{OQMD}, etc.). In recent years, computational methods driven by Machine Learning (ML) have proven effective in accelerating the exploration of the chemical space, assisting in the identification of dielectric materials \citep{riebesell2024pushing}, nickel-based superalloys \citep{ml_nickel_superallo} and superhard materials \citep{ml_superhard}. Despite the broad perspectives opened up by data-driven methods, the horizon of available properties to leverage ML towards the discovery of specific material classes is still quite narrow due to the scarcity and dispersity of available data to train ML models. 
Many data-driven approaches are based on computed data and thus subject to the approximations and limitations of the calculation themselves. Experimental data are generally not available at scale.
Industrial applications frequently require \textit{exceptional} compounds \citep{schrier_exceptional}, often exhibiting a  counterintuitive combination of two or more chemical properties. This poses significant challenges to current data-driven frameworks, as conventional material databases may lack the necessary information to effectively guide ML in discovering materials tailored at specific applications.

Transparent conducting materials (TCMs) fully exemplify the category of exceptional compounds. These represent a class of semiconductors showing simultaneously high electrical conductivity, and low absorption in the range of visible light. This unique behaviour is often enforced in practice by a process known as \emph{doping}, where \modif{additional components} are introduced into an intrinsic semiconductor to modulate its optoelectronic properties. 
Conventional transparent conductors are typically achieved by doping metal oxide semiconductors like $\text{In}_{2}\text{O}_{\text{3}}$, $\text{Sn}\text{O}_\text{2}$, $\text{CdO}$ and $\text{ZnO}$. Among various classes of TCMs, tin-doped indium oxide (ITO) stands out as the most common one typically used in high value applications such as displays due to the scarcity of indium, while fluorine-doped tin oxide (FTO) has been widely adopted in larger area applications such as solar control glazing and transparent electrodes for solar cells \citep{most_used_tco2}. Although the existing set of TCMs currently addresses the demands imposed by modern optoelectronic applications, the scarcity of raw materials, together with the high costs of vapour deposition techniques, drive researchers to look for alternative solutions \citep{recent_tcm_photo, tcms_optoelectronics}.
Previous literature using ML in the TCMs field has investigated the optimization of existing semiconductors \cite{woodsRob} , or focused on well-defined phase-fields \cite{Xiong} \cite{Sutton}, and progress has been hindered due to the absence of adequate datasets of experimental optoelectronic properties.

\noindent In this work, we propose a data-driven framework to accelerate the discovery of new TCMs.
To address the shortage of available data, we create and validate databases of chemical formulas reporting experimental room-temperature conductivity and band gap measurements. We utilize the obtained data to train state-of-the-art (SOTA) ML models that leverage the stoichiometry of input materials, given the typical absence of structural information in materials discovery tasks. Furthermore, we assess the performance of trained models using a custom evaluation framework, designed to determine whether ML can identify previously unseen classes of TCMs. To test the proposed framework, we further utilize a list of 55 experimentally-reported chemical compositions sourced from entries across MPDS \cite{mpds}, Pearson \cite{pearson}, and ICSD databases \cite{icsd}. 
We use this list to empirically demonstrate the effectiveness of ML in accelerating the identification of new materials that are likely to display TCMs characteristics.
The main contributions of this study can be summarized as follows:
\begin{itemize}
	\item We create \modif{two datasets} of experimentally-reported optoelectronic properties, 1) a dataset of electrical conductivity is collated and curated from data residing in the MPDS and 2) we augment a published band gap dataset. Both datasets serve as a foundation for training ML models aimed at the identification of TCMs.

 	\item We evaluate SOTA ML models for property-prediction on the proposed experimental datasets. 

	\item We empirically measure the ability of ML models to identify new classes of TCMs through a bespoke evaluation method.

    \item We compile a list of 55 compositions across various databases and we empirically demonstrate the potential of ML in accelerating the identification of materials that are likely to exhibit TCMs characteristics.
\end{itemize}
\section{Related work}
\paragraph{Computationally-guided search for new TCMs}
DFT has primarily enabled a computational exploration of various material classes, including TCMs.
Notably, \citet{woodsRob} curated an experimental dataset comprising 74 bulk structures of well-known TCMs with the goal of computing a set of DFT-based descriptors that would capture essential features of these materials for computational screening purposes. \citet{hautier_mass} employed a  high-throughput computational approach to identify oxides with low electron effective mass. They also assessed the band gap of the most promising candidates and proposed potentially novel n-type transparent conducting oxides. The increasing accessibility of materials data has also facilitated data-driven frameworks for ML-guided search for new materials. \citet{tcms_ml} conducted a study that explored the application of ML to predict new TCMs. They utilized data on formation energy and band gap obtained from a Kaggle competition focused on TCMs discovery \citep{tcms_ml}. 
Despite the promises established by computational modelling,
challenges such as high computational cost and systematic errors in DFT-based approaches, along with the scarcity of suitable datasets in the realm of ML, have posed important obstacles to the search for new such materials. 
\paragraph{Data-driven identification of optoelectronic properties} Electronic transport and optical data on semiconductors have been gathered and evaluated in the context of thermoelectrics~\citep{ml_thermo, ucsb} and of band gap~\citep{zhuo_2018, wang_2022}.  Studies have then evaluated different ML approaches in combination with data extracted from the \modif{University of California Santa Barbara (UCSB)} dataset to predict the electrical conductivity of materials \cite{mukherjee, na2022dop}. Furthermore, DFT-calculated datasets for electron transport properties have also been proposed~\citep{Ricci, jarvis_dft, yao, miyazaki} and utilized for different tasks ranging from data visualization, to ML property prediction. The availability of experimental datasets has remained rather limited~\citep{katsura, priya, ucsb, public_thermo} , with most available datasets reaching the order of $\sim 10^2$ entries. Furthermore, experimental data often encompass minimal chemical diversity, primarily due to the difficulties in obtaining reliable measurements. 
 These two crucial issues (limited datasets size and narrow chemical diversity) heavily limit the application of data-driven methods for the prediction of electronic properties. In the case of band gap, the extensive availability of entries derived from DFT calculations~\citep{matproj, aflow, OQMD} has, in part, mitigated the problem of data scarcity, specifically because this property is more feasible to theoretical simulations compared to electron transport properties. \modif{However, significant challenges persist in the prediction of experimental band gaps due to the underestimation of band gaps calculated using the high-throughput DFT approaches of large databases~\citep{cohen2008dft}} and imbalance between metals and non-metals in the available datasets~\citep{riebesell2024pushing}.
\section{Databases overview}\label{sec:datasets}
A well-established figure of merit for TCMs can be identified as the ratio of electrical conductivity $(\sigma)$ to the optical absorption coefficient $(\alpha)$ \cite{gordon2000}:
\begin{equation}\label{fom}
	\varphi_{TCM} = \frac{\sigma}{\alpha} \, .
\end{equation}

A well-performing TCM should combine high electrical conductivity with low absorption of visible light. 
Therefore, to accomodate $\varphi_{TCM}$ within a data-driven perspective, it would be necessary to rely on abundance of data in terms of $\sigma$ and $\alpha$.
Typically, datasets containing these properties are scarce and fragmented across numerous sources in the literature. 
To address the limitation of optical property data, we adopt the band gap ($E_g$) as a proxy for optical transparency, motivated by the abundance of this information in the existing literature \citep{matproj, OQMD}. The band gap is a crucial parameter that influences materials' optical properties. 
A material with a band gap exceeding the energy of visible light (approximately 3 eV) appears generally transparent, as photons within this range lack the energy to excite electrons across the band gap. Thus, by choosing materials with band gaps greater than 3 eV, we can identify materials that are likely to
exhibit transparency in the visible spectrum.
To enable a ML approach, we have created and validated two experimental datasets of room-temperature conductivity and band gap measurements, to be used as foundation for training SOTA ML models for the discovery of new TCMs. Below, we detail the creation of these databases, a key contribution of this work. \modif{Both datasets were tailored to remove unphysical entries by expert assessment and to ensure that a wide range of chemistries were included, resulting in datasets well-balanced between metals and non-metals as discussed below.}
\paragraph{Electrical conductivity dataset}
The electrical conductivity dataset was constructed using two primary data sources. Initially, conductivity and resistivity data were gathered from the Materials Platform for Data Science (MPDS) \cite{mpds} (38,068 entries). This source was supplemented with the UCSB dataset \cite{ucsb} (1,794 entries), which provides a range of experimental thermoelectric properties, including electrical conductivity. In total, we compiled a raw dataset comprising $39,862$ material entries with associated conductivity measurements at various temperatures.
Several preprocessing steps were conducted on the raw data. Initially, we excluded all pure elements and noble gases and selected all chemical formulas reported within a window of room temperature ($298 \pm 5$ K), reducing the dataset to 14,307 entries. Given the experimental nature of utilized data, it is common to encounter several material entries where different measurements are documented for identical chemical formulas at the same temperatures. This variance is inherently linked to the different experimental conditions under which these measurements were conducted. To process raw data in view of statistical estimation, we initially considered the distributions of measurements corresponding to duplicated chemical formulas, discarding those groups associated to a standard deviation exceeding 10 S/cm. 
\begin{figure*}[t]
	\includegraphics[width=0.508\textwidth]{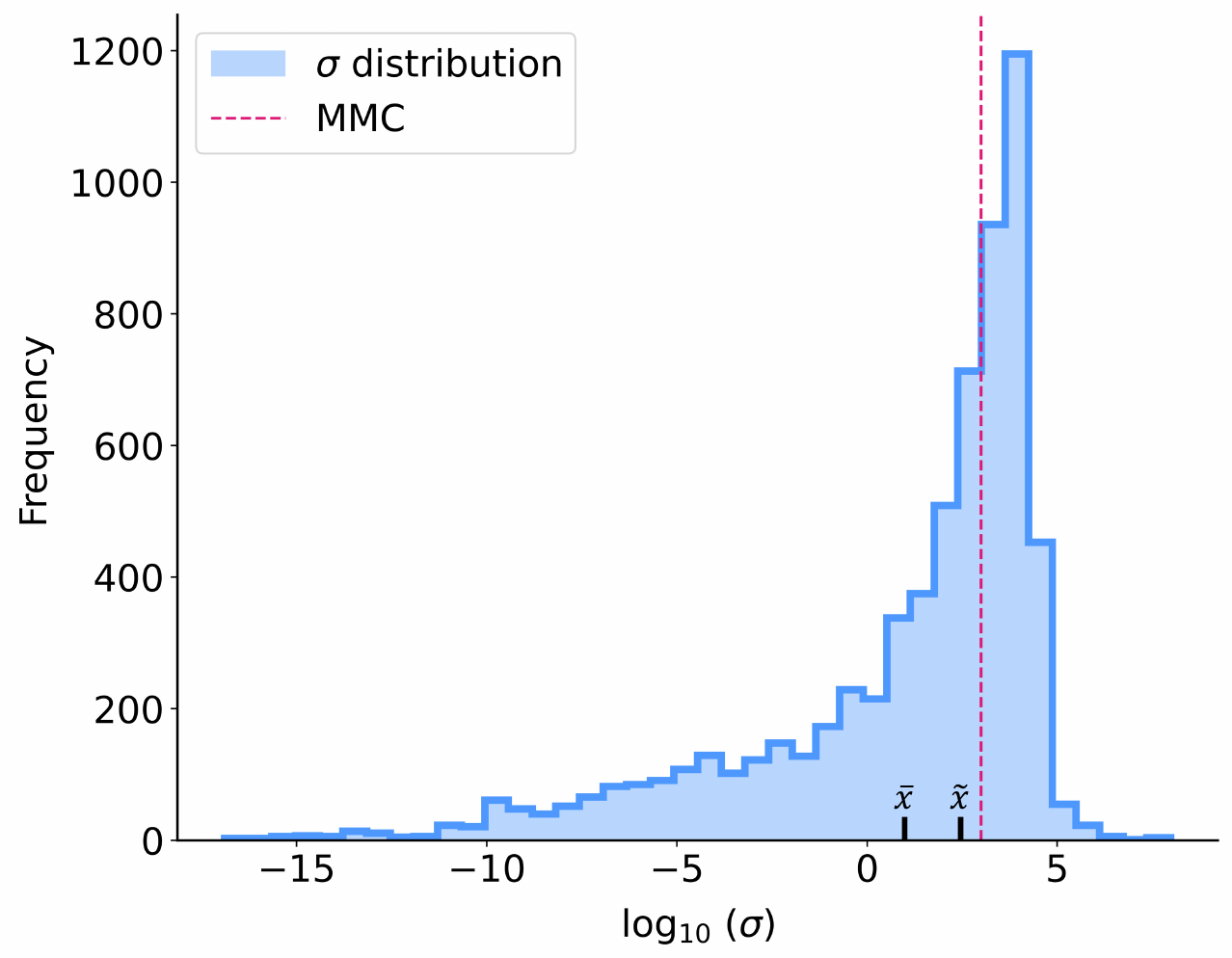}
	\includegraphics[width=0.48\textwidth]{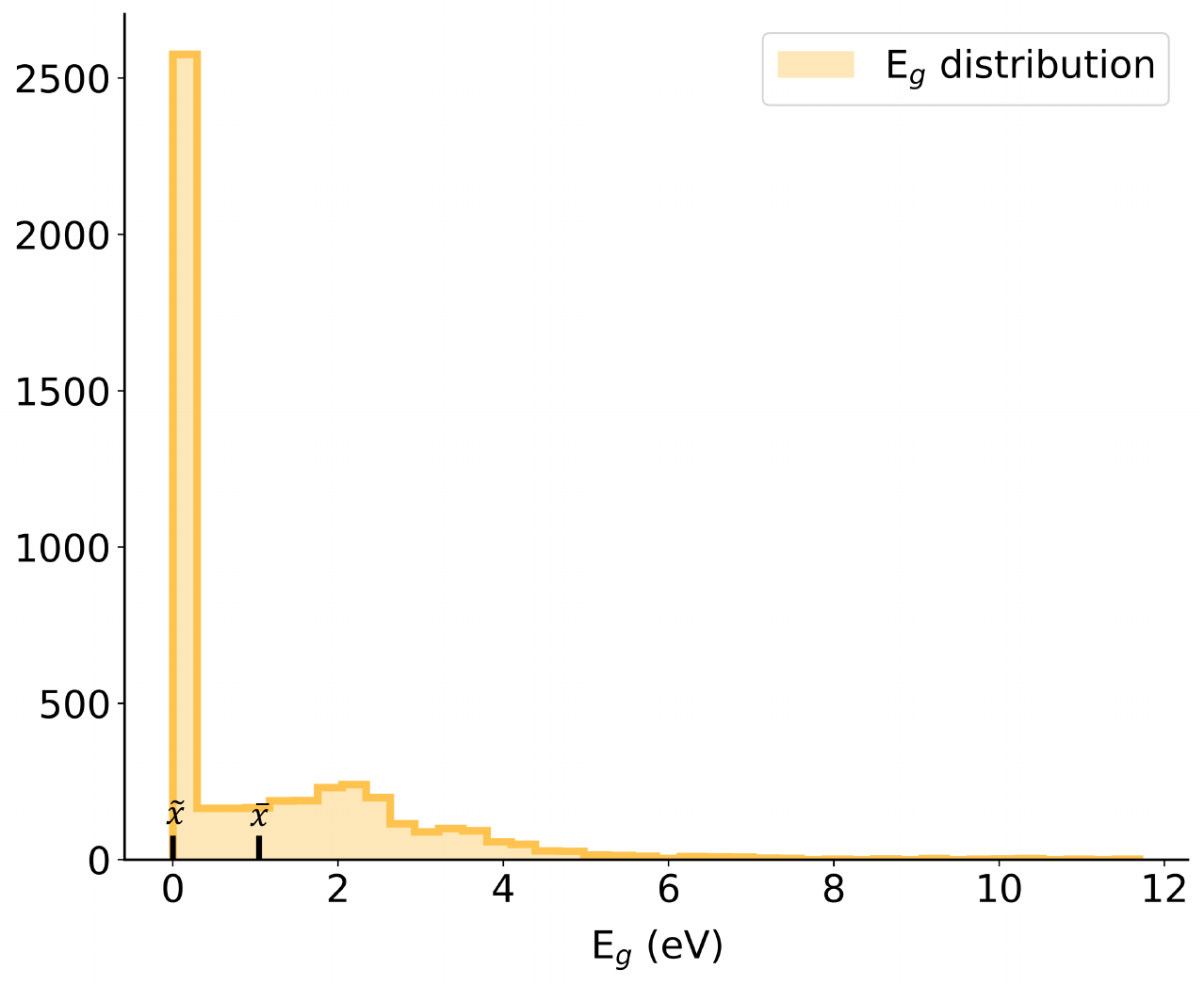}
	\caption{Data distributions for $\sigma$ (left) and $E_g$ (right). $\bar{x}$ and $\tilde{x}$ denote the \textit{mean} and the \textit{median}, respectively. The purple dotted line on $\sigma$ distribution indicates the minimum metallic conductivity $\sigma_{min}=10^3 \, \text{(S/cm)}$.}
	\label{fig:datasets}
\end{figure*}
Furthermore, we excluded entries with conductivity measurements falling outside of 4 standard deviations from the mean, resulting in a processed dataset containing 6,503 material entries. At this stage, we performed a meticulous validation, which involved a line-by-line review of the obtained data by domain experts, referring back to the original literature on suspicious entries, to ensure the accuracy of the reported conductivity measurements, alongside the correctness of the corresponding chemical formulas.
To facilitate the validation process, automated nonsense-detection strategies were implemented to systematically identify anomalous conductivity measurements associated to the reported material entries. This involved inferring the oxidation states of the chemical elements in each composition, to ensure the feasibility of different chemical species, in accordance with their corresponding conductivity measurements. 
First, Comgen \citep{clymo2024comgen} was used to infer the oxidation states of chemical elements in each composition. These were used to verify the feasibility of the chemical species in the composition, in accordance with the reported conductivity measurement.
For example, closed-shell oxides are expected to exhibit low conductivities. Therefore, reported entries corresponding to closed-shell oxides with a conductivity higher than a threshold set to $10^{-6} \, \text{S/cm}$ were automatically flagged by the nonsense-detection tool \modif{for further expert consideration}. Additionally, we incorporated experimental conductivities for several chemical families that were absent, such as the alkaline earth oxides, binary and ternary oxides \modif{including materials selected to represent each integer transition metal oxidation state as far as available data allow, as well as known TCMs (reported in Table~\ref{table:tcms}).} We end up with a final, validated database comprising 6,592 material entries, with a mean $\bar{x}$ of  $0.99 \, \log_{10}$ (S/cm), a median $\tilde{x}$ of $2.46 \, \log_{10}$ S/cm and an interquartile range ($50\%$ of data; materials from the 25th to the 75th percentile of $\log_{10}(\sigma)$) spanning from $-0.44$ to $3.67$ $\log_{10}$ S/cm. The data distribution of conductivity dataset is shown on the left of Figure~\ref{fig:datasets} .
To understand the distribution of metals and non-metals in our conductivity dataset, we utilize the theoretical notion of minimum metallic conductivity (MMC), as introduced by \citet{mott1985metal}. This indicates a threshold below which materials exhibit semiconductor-like behavior.
Thus, compounds with conductivity above this threshold display metallic characteristics, while those below it show a non-metallic behavior. For our analysis, we adopt a threshold value of $\sigma_{min} = 10^3$ S/cm, represented by the purple dotted line in Figure~\ref{fig:datasets} (left), which has been experimentally observed for many transition metal compounds near the metal-insulator transition \citep{chudnovskii1978metal}. Applying this criterion, we identified 2,675 metals in the dataset ($ \approx 41\%$), and 3,917 materials ($\approx 59\%$) exhibiting non-metallic conductivity.
\paragraph{Band gap dataset} The initial band gap data was sourced from a well-known experimental dataset proposed by \citet{zhuo_2018}.
The original dataset comprises $6,354$ material entries with experimental band gap measurements determined from optical and transport measurements. Preprocessing steps were applied to the raw data. Specifically, we excluded groups of duplicated formulas with band gap measurements having a standard deviation greater than 0.1 eV.  This preprocessing approach is similar to the one used for creating the \verb|matbench_expt_gap| dataset, available on the Matbench platform~\citep{matbench}.  All the entries associated with noble gases and pure elements have been discarded. Additionally, entries with band gap measurements exceeding 4 standard deviations from the mean have been excluded, leading to a processed dataset of 4,732 material entries. As in the case of conductivity, the obtained pool of data has been expanded by including experimental band gap measurements of binary and ternary oxides not already in the dataset, along with known TCMs, reported in Table~\ref{table:tcms}. These preprocessing steps resulted in a final dataset comprising 4,767 material entries, with a mean $\bar{x}$ of $1.04$ (eV), a median $\tilde{x}$ of $0.00$ (eV), and an interquartile range spanning from $0.00$ to $1.93$ eV . The data distribution of the band gap dataset is shown on the right in Figure~\ref{fig:datasets} . We observe a balanced representation of metals ($E_g = 0$) and non-metals ($E_g > 0$) in the created dataset. The group of metals comprises 2,426 material entries ($ \approx 51\%$), while non-metals encompass 2,341 entries ($ \approx 49\%$).
\setlength{\tabcolsep}{23pt}
\renewcommand{\arraystretch}{1.3} 
\begin{table}[t]
	\vskip 0.15in
	\begin{center}
		\begin{tabular}{lccc }
			\toprule
			\textbf{TCMs family} & $N$  & $\sigma$ ($\log_{10}$ S/cm) ($\mu \pm s$) & $E_g$ (eV) ($\mu \pm s$)\\
			\midrule
			SnO$_{2}$: Ga \citep{sno2:ga} & 3 &$2.52 \pm 0.03$ & $3.77 \pm 0.03$\\
			SnO$_{2}$: In \citep{sno2:in}& 4 & $2.26 \pm 0.75$&$3.83 \pm 0.09$\\
			SnO$_{2}$: Mn \citep{sno2:mn} & 3 & $2.07 \pm 0.01$ & $4.07 \pm 0.03$ \\
			SnO$_{2}$: Ta \citep{sno2:ta} & 4 & $2.52 \pm 0.54$& $4.16 \pm 0.11$ \\
			SnO$_{2}$: Ti \citep{sno2:ti} & 5 & $2.73 \pm 0.06$ & $3.80 \pm 0.06$ \\
			SnO$_{2}$: W \citep{sno2:w} & 4 & $2.23\pm 0.22$& $4.23 \pm 0.68$ \\
			In$_2$O$_3$: Sn  \citep{in2o3_sn_1, in2o3_sn_2, in2o3_sn_3} (ITO) & 3 & $2.65 \pm 0.64$ & $3.73\pm 0.29$\\
			ZnO: Al-Sn \citep{zno:al-sn} & 4 & $2.58 \pm 0.18$ & $3.80 \pm 0.16$\\
			ZnO: Al \cite{zno:al} & 3 & $3.43 \pm 0.57$ & $3.61 \pm 0.05$\\
			ZnO: Ga \cite{zno:ga} & 6 & $3.93\pm 0.29$ & $3.64 \pm 0.05$\\
			\bottomrule
		\end{tabular}
	\end{center}
	\caption{Various families of TCMs, each with distinct $N$ representatives associated to a specific doping level (at\%). We report the mean ($\mu$) and standard deviation ($s$) related to conductivity and band gap measurements for different families.}
	\label{table:tcms}
\end{table}

\section{Methods}
In this section, we introduce both the ML models and the evaluation methods considered in this study.
\subsection{Models}
\paragraph{Random forest (RF)}\cite{random_forest} A classic ML approach that is well established in the field of materials informatics and has been applied in a variety of tasks, from predicting band gap energy \cite{rf_gap_cbfv} to identifying thermoelectric  and mechanical properties \citep{riebesell_thermoelectric,rf_mech_strength}. The algorithm involves a combination of various weak learners that are trained on resampled versions of the original dataset and with different subsets of features. This has the effect of reducing model variance by decorrelating individual decision trees. In practice, it is commonly used in tandem with materials representations obtained by aggregating attributes from individual elements of the periodic table. These features are typically denominated structure or composition-based feature vectors, given that they are obtained using the stoichiometry alone \citep{magpie}, or other known attributes from the underlying crystalline structure \cite{cryst_descriptors}.

\paragraph{CrabNet} \cite{crabnet} A neural-network architecture based on the paradigm established by transformers \citep{transformers}. The core idea of these models relies on self-attention, which finds an early application in the field of natural language processing: intuitively, given a sequence (phrase) of $n$ tokens $\bm{x}_1, \bm{x}_2, \dots , \bm{x}_n$, the goal is to learn new, context-aware representations $\bm{y}_1, \bm{y}_2 , \dots , \bm{y}_N$, with a richer semantic structure. This is achieved by learning attention scores between word pairs within the phrase.
\noindent In the context of materials science, the input tokens can be viewed as elements of a chemical composition. Attention scores, computed via self-attention, can then be utilized to adjust the overall material representation for predicting a specific property of interest.
CrabNet has delivered remarkable outcomes in predicting chemical and physical properties of materials when only the composition is available \cite{matbench}. It frequently serves as a SOTA model in scenarios where property predictions are solely reliant on the chemical composition of materials \cite{discover, hargreaves_2023, cratenet, lee2023clcs}. For further details regarding the underlying architecture, we refer to the original paper \cite{crabnet}. 
\subsection{Evaluation}\label{subsec:eval}
In our goal of identifying the constitutive properties of the materials of interest, we stay aligned to previous work \citep{wang_2022, mukherjee_sigma, nguyen_sigma, hargreaves_2023} and adopt a regression task.  In this context, the goal is to train ML models to predict numerical values associated to the corresponding material properties. 
It is worth mentioning that a classification task may be considered too, directly determining whether the predicted material meets the specified criteria or not and thus falls into the category of TCMs. However, we argue that adopting a classification approach in this context might sacrifice valuable interpretability. Rather than simply classifying materials as TCMs or non-TCMs, regression models provide continuous numerical predictions for properties like conductivity and band gap. This granularity offers a more precise understanding of each material’s performance, allowing us to evaluate how close each material is to meeting the TCM criteria. 
To assess the performance of trained ML models, we utilize different evaluation schemes: K-fold, a conventional method deeply rooted in statistical learning theory \citep{tibshirani_2001}, is commonly employed; additionally, Leave-One-Cluster-Out Cross-Validation (LOCO-CV) \cite{meredig_2018} stands as an alternative method targeting the assessment of chemical extrapolation, crucial for discovering new materials, absent in the training data. Furthermore, we introduce a third evaluation method designed to offer nuanced interpretability within the task at hand, namely the discovery of novel TCMs. Details outlining each of these methods are provided in the following.
\paragraph{K-fold} Validation process involves quantifying the deviation between predictions and real underlying targets, in a portion of the dataset that is held out at training stage. 
This is typically achieved with a K-fold cross validation, which consists in splitting the original dataset in $k$ equally-sized folds ($k=5$ in this study), and in turn, training the model on $k-1$ of these and using the remaining one for evaluation, to have an estimate of the average test error. While K-fold cross-validation is a well-established and commonly used procedure for assessing the performance of ML models, it may not serve as an accurate indicator of their extrapolation capability in the context of materials discovery. The main concern arises from the fact that within a K-fold approach, similar stoichiometries can end up in both training and test data. As a consequence, the model might be provided with a relatively favorable scenario, where it can effortlessly interpolate between known stoichiometries, rather than being truly challenged to extrapolate beyond the observed data. This aspect is intrinsically connected to the redundancy of material datasets \citep{ottomano, critical_ex_gen}, which inevitably leads to overestimating the performance of ML models \citep{omee2024structurebased}, unless bespoke evaluation schemes are designed to quantify the extrapolation error. This phenomenon can potentially mask any limitations or weaknesses in the models' ability to generalize to new and unseen materials, undermining the overall predictive power in the context of materials discovery. 
\begin{figure*}[t]
	\includegraphics[width=0.98\textwidth]{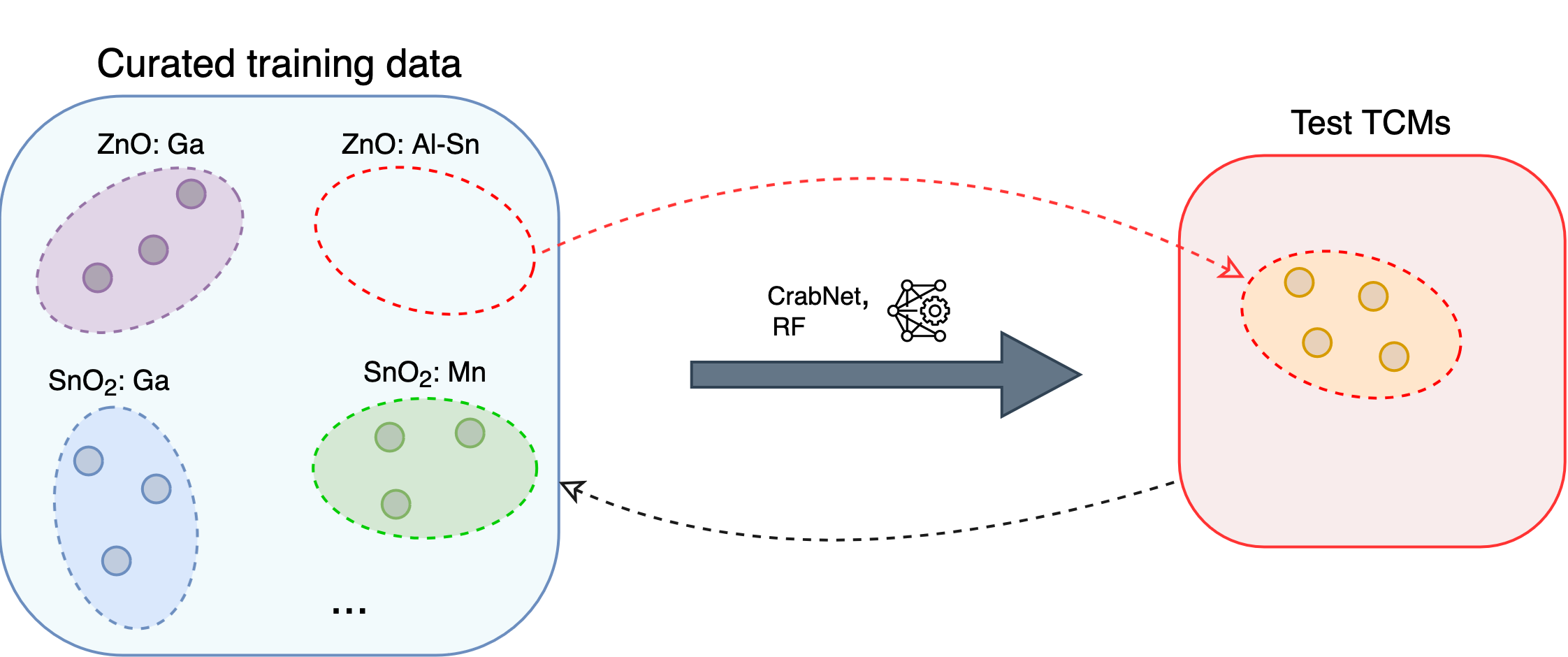}
	\caption{Schematic representation of the proposed evaluation to simulate the discovery of new TCMs: following an iterative scheme, a specific family of known TCMs is placed in the test set, while ML models are trained on the remaining TCMs within training data. This procedure repeats for each available TCM family.}
	\label{fig:tasks}
\end{figure*}

\paragraph{LOCO-CV}While K-fold cross-validation remains valuable for assessing models' performance within the training distribution, it may not fully capture the crucial aspect of extrapolation in any materials discovery task involving ML. In addition to K-fold, we employ a LOCO-CV \cite{meredig_2018} evaluation scheme . With LOCO, the folds are not randomly generated, but rather constructed by grouping together material families that exhibit chemical similarity. This method provides a more refined evaluation of models' performance by focusing on the ability to generalize to new material groups. For example, one might be interested in assessing the extrapolation power of a ML model in predicting a group of oxides given that this family was unobserved at training stage. Different techniques can be employed to effectively implement this approach: in general, when featurizing input chemical formulas, the initial step often involves employing the K-means algorithm \cite{K-means} to generate a predetermined number of distinct clusters.
However, a challenge arises due to the eventual \textit{disparity} in the sizes of material groups, which can introduce excessive variance during the evaluation process. To address this scenario, prior observations have indicated that applying kernel functions to the material representations can promote more equitable cluster sizes and enhance the invariance of the resulting clusters with respect to the chosen representation for the input chemical formulas \cite{random_proj_sam}. Kernels are mathematical functions that transform the input data into a higher-dimensional feature space where better linear separability is possible \cite{kernels}. We employ a radial basis function (RBF) \cite{rbf} to process featurized chemical formulas before applying the K-means algorithm.

\paragraph{Leave-one-TCM-family-out} In principle, LOCO-CV can be considered as a well-motivated method to evaluate the chemical extrapolation of ML models under consideration. However, the assessment is often limited by the varying sizes of material clusters, which lead to a noisy evaluation and to an increased variance in the assessed metrics. Moreover, it is common for the data folds generated within a LOCO-CV setting to result from the sequential application of various algorithms, which in turn leads to a limited interpretability regarding the resulting material clusters. To gather empirical evidence regarding the ability of ML to uncover novel compounds for real-world applications, \modif{we propose a new evaluation strategy that we denote as \emph{leave-one-TCM-family-out}}. This evaluation method aims at providing empirical evidence on whether ML can discover new TCMs, given prior knowledge from known materials.
For a comprehensive analysis, we initially gather diverse families of established TCM materials.  In Table \ref{table:tcms} we present a summary of different material families examined in this study, along with the count of associated representatives and the average values of reported electrical conductivity and band gap measurements. In total, we have compiled 39 examples of established TCMs from the existing literature.
Different representatives within the same family reflect different concentrations (at\%) of the corresponding dopant element. Drawing insights from the statistics of reported TCMs and from prior scientific knowledge, we establish an identification criterion aimed at understanding whether ML can successfully identify TCM materials: specifically, a TCM will be successfully identified if the corresponding predictions for electrical conductivity and band gap exceed $10^2  \, \text{S/cm}$ and $3$ eV, respectively. Intuitively, we want to investigate whether ML models can discriminate the behavior of doped semiconductors, and detect a significant level of electrical conductivity, even in situations where there exists a non-negligible band gap.
In the leave-one-TCM-family-out evaluation scheme, we exclude a specific family of TCMs from the training set, while retaining other representative materials.
This assessment seeks to offer empirical evidence about the ability of ML to uncover novel material families, leveraging the existing knowledge as a starting point. In practice, we are asking ML models to identify new stoichiometric combinations in the test set previously unobserved at training stage. If one of the TCM families, either $\text{SnO}_2\text{: In}$ or $\text{In}_2\text{O}_3\text{: Sn}$, is present in the test set, the other is excluded from training, as they share the same chemical elements, despite representing two different sets of TCMs.
To quantify the success rate in the proposed evaluation, we establish a new metric named \textit{family-discovery-rate} (FDR), which considers the percentage of discovered TCMs families by ML, with success defined as the accurate prediction of at least one representative from the overall family, when that family is removed from the training data. We define it as:
\begin{equation}
	FDR(\%) := \frac{N^*_f}{N_f} \times 100 \, ,
\end{equation}
where $N_f$ represents the total number of families and $N^*_f$ is the count of correctly predicted families.
In Figure \ref{fig:tasks} , we provide a visual overview of the proposed evaluation scheme.
 \begin{figure*}[t]
	\includegraphics[width=0.48\textwidth]{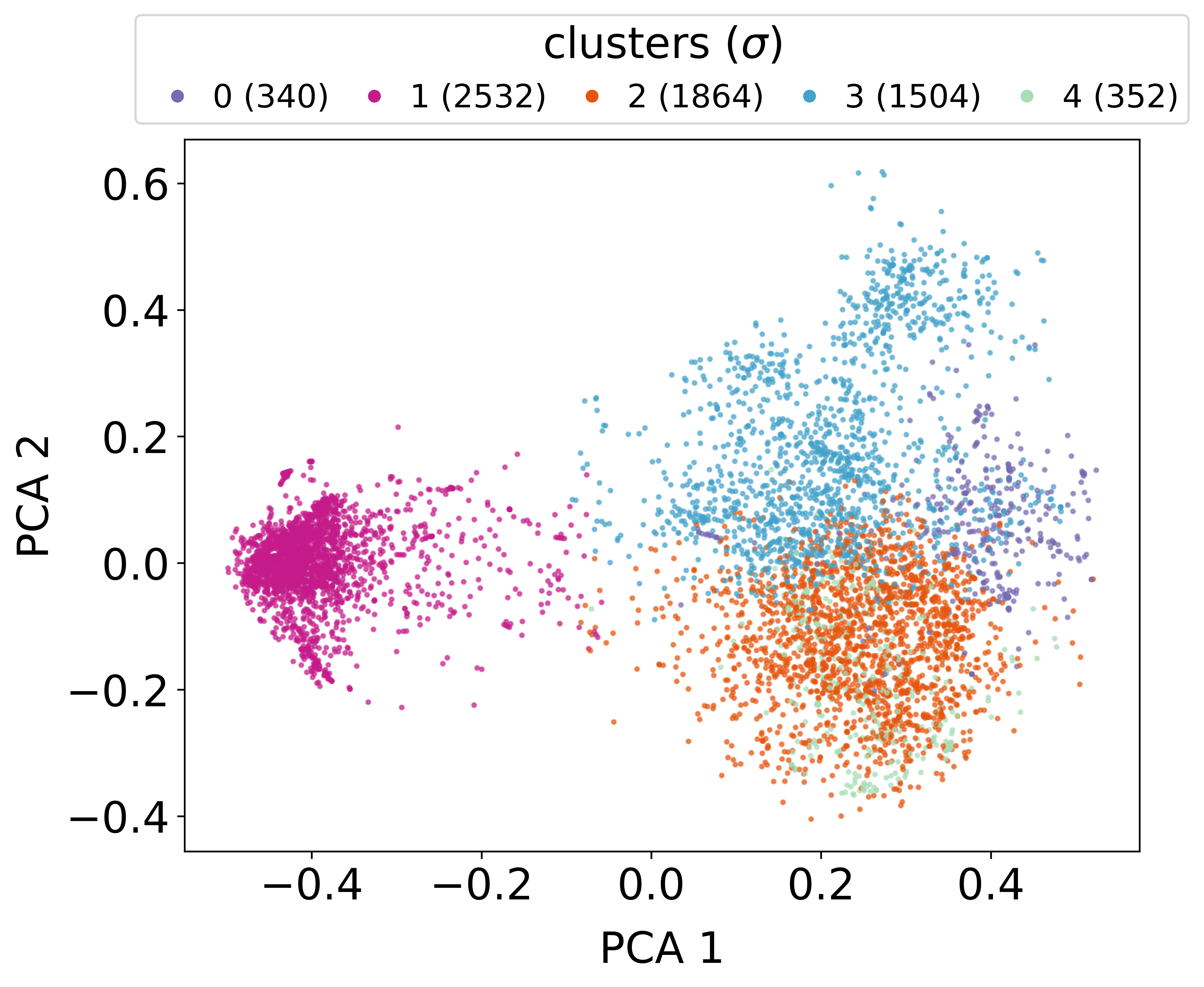}
	\includegraphics[width=0.48\textwidth]{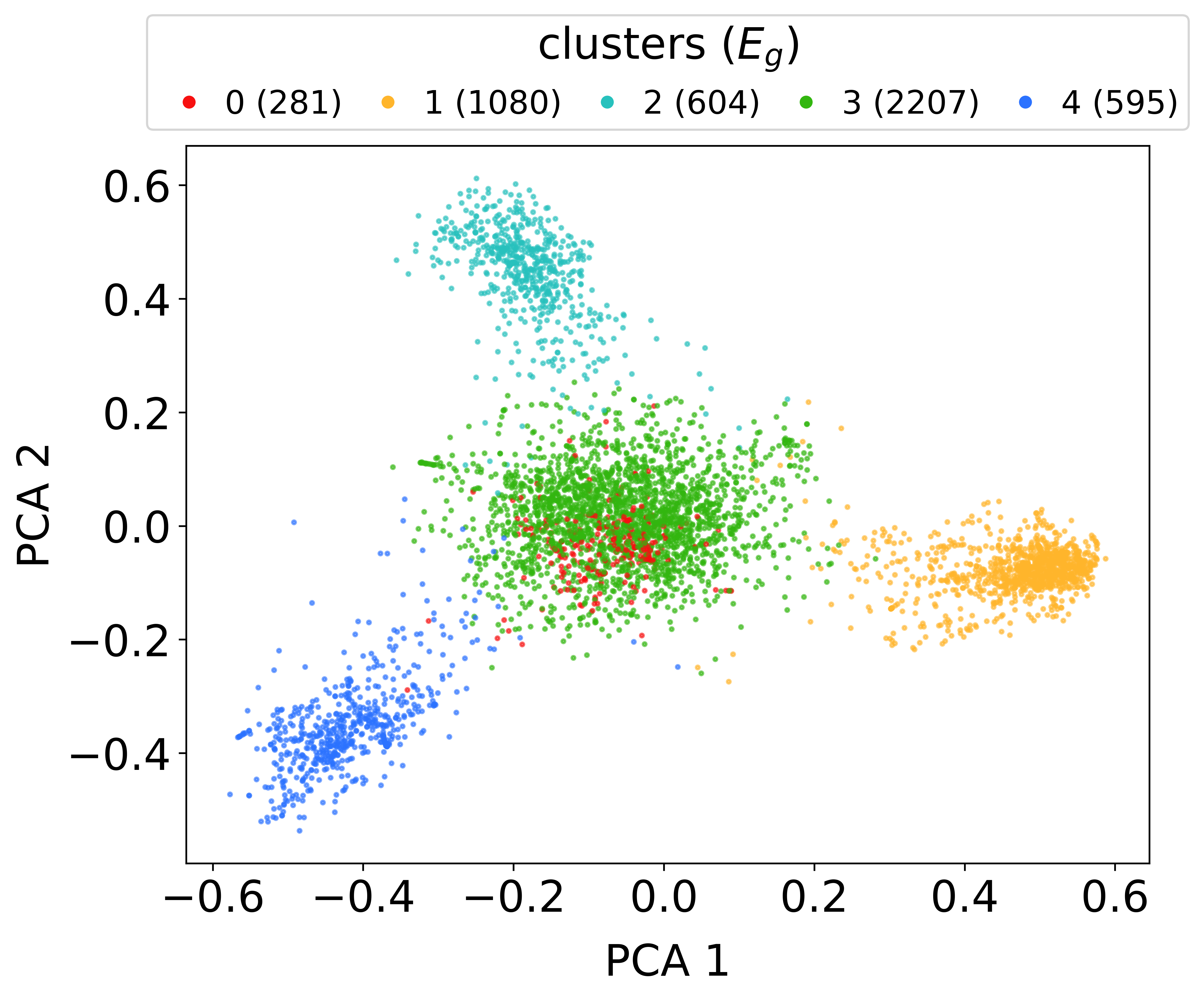}
	\caption{LOCO-CV material clusters obtained separately for the conductivity dataset (left) and for the band gap dataset (right).}
	\label{fig:loco}
\end{figure*}
\FloatBarrier
\setlength{\tabcolsep}{18pt}
\renewcommand{\arraystretch}{1.3} 
\begin{table*}[!t]
	\centering
	\caption{ML models evaluation for electrical conductivity ($\sigma$) prediction (log$_{10}$ (S/cm)). Best-performing results are shown in green, while second best-performing are shown in yellow, when there is an overlap in the uncertainty bands. Upward and downward arrows indicate the desired direction for improvement for the corresponding metric.}
	\vspace{1mm}
	\resizebox{.95\textwidth}{!}{
		\begin{tabular}{ccccc}
			\hline
			\addlinespace
			\multirow{2}{*}{\textbf{Model}} &
			\multicolumn{2}{c}{\textbf{KFold}} &
			\multicolumn{2}{c}{\textbf{LOCO-CV}}\\
			& {MAE $\downarrow$} & {$R^2$ $\uparrow$} & {MAE $\downarrow$} & {$R^2$ $\uparrow$}\\
			\hline
			\addlinespace
			RF + Magpie & \cellcolor{green!20} $\bm{1.27\pm0.03}$ &  \cellcolor{green!20} $\bm{0.71 \pm 0.02}$ & \cellcolor{yellow!20}$2.11 \pm 1.00$ &\cellcolor{yellow!20}  $0.08 \pm 0.26$ \\
			CrabNet & \cellcolor{yellow!20} $1.30\pm 0.04$ & $0.67 \pm 0.02$ & \cellcolor{green!20} $\bm{1.81 \pm 0.87}$ & \cellcolor{yellow!20}$ 0.13 \pm 0.22$\\
			\bottomrule
		\end{tabular}
	}
	\label{table:sigma}
\end{table*}
\begin{table*}[!t]
	\centering
	\caption{ML models evaluation for band gap ($E_g$) prediction (eV). Best-performing results are shown in green, while second best-performing are shown in yellow, when there is an overlap in the uncertainty bands. '/' indicates a negative $R^2$ score, and thus the failure of the corresponding regression task. Upward and downward arrows indicate the desired direction for improvement for the corresponding metric.}
	\vspace{2mm}
	\resizebox{.95\textwidth}{!}{
		\begin{tabular}{ccccc}
			\hline 
			\addlinespace
			\multirow{2}{*}{\textbf{Model}} &
			\multicolumn{2}{c}{\textbf{KFold}} &
			\multicolumn{2}{c}{\textbf{LOCO-CV}}\\
			& {MAE $\downarrow$} & {$R^2$ $\uparrow$} & {MAE $\downarrow$} & {$R^2$ $\uparrow$}\\
			\hline
			\addlinespace
			RF + Magpie & $0.41 \pm 0.02$ &  \cellcolor{yellow!20} $0.70 \pm 0.03$ & \cellcolor{yellow!20} $0.86 \pm 0.40$ & $/$ \\
			CrabNet &  \cellcolor{green!20} $\bm{0.30 \pm 0.02}$ &  \cellcolor{green!20} $\bm{0.73 \pm 0.05}$ & \cellcolor{green!20} $\bm{0.56 \pm 0.32}$ &\cellcolor{green!20} $\bm{0.47 \pm 0.15}$\\
			\bottomrule
		\end{tabular}
	}
	\label{table:gap}
\end{table*}
\FloatBarrier
\section{Results}\label{sec:results}
Since the primary task can be formulated as a regression problem, we utilize mean absolute error (MAE) and coefficient of determination ($R^2$) as evaluation metrics to assess models' performance. In our preliminary analysis, we assessed an additional architecture, DopNet \citep{na2022dop}, which was originally proposed to predict various thermoelectric properties by taking into account dopants' influence. We have been motivated in experimenting with this model by the prevalent nature of doped semiconductors in the pool of known TCMs. 
However, our analysis demonstrated a lower performance compared to CrabNet and RF ($ \approx 40\%$ worse in terms of MAE for conductivity prediction, compared to CrabNet). This can be attributed to the use of a non-optimized version of the architecture, where attributes such as the threshold for distinguishing dopants from the host, and the maximum number of dopants per composition are hyperparameters. We defer a more precise investigation of this architecture to future work. For band gap prediction, CrabNet undergoes pre-training on a dataset of DFT-computed band gaps sourced from the Materials Project \citep{matproj}. \modif{This pre-trained model} is then fine-tuned on the curated experimental band gap dataset (results for CrabNet’s band gap predictions, shown in Table \ref{table:gap}, pertain to this fine-tuned model). We adopt a transfer learning approach to mitigate well-known ML limitations in band gap prediction, which often result in several metallic materials erroneously identified as semiconductors or insulators \citep{riebesell2024pushing}.
\subsection{KFold \& LOCO-CV}\label{subsec:kfold_loco}
\begin{figure*}[t]
	\centering
	\includegraphics[width=0.90\textwidth]{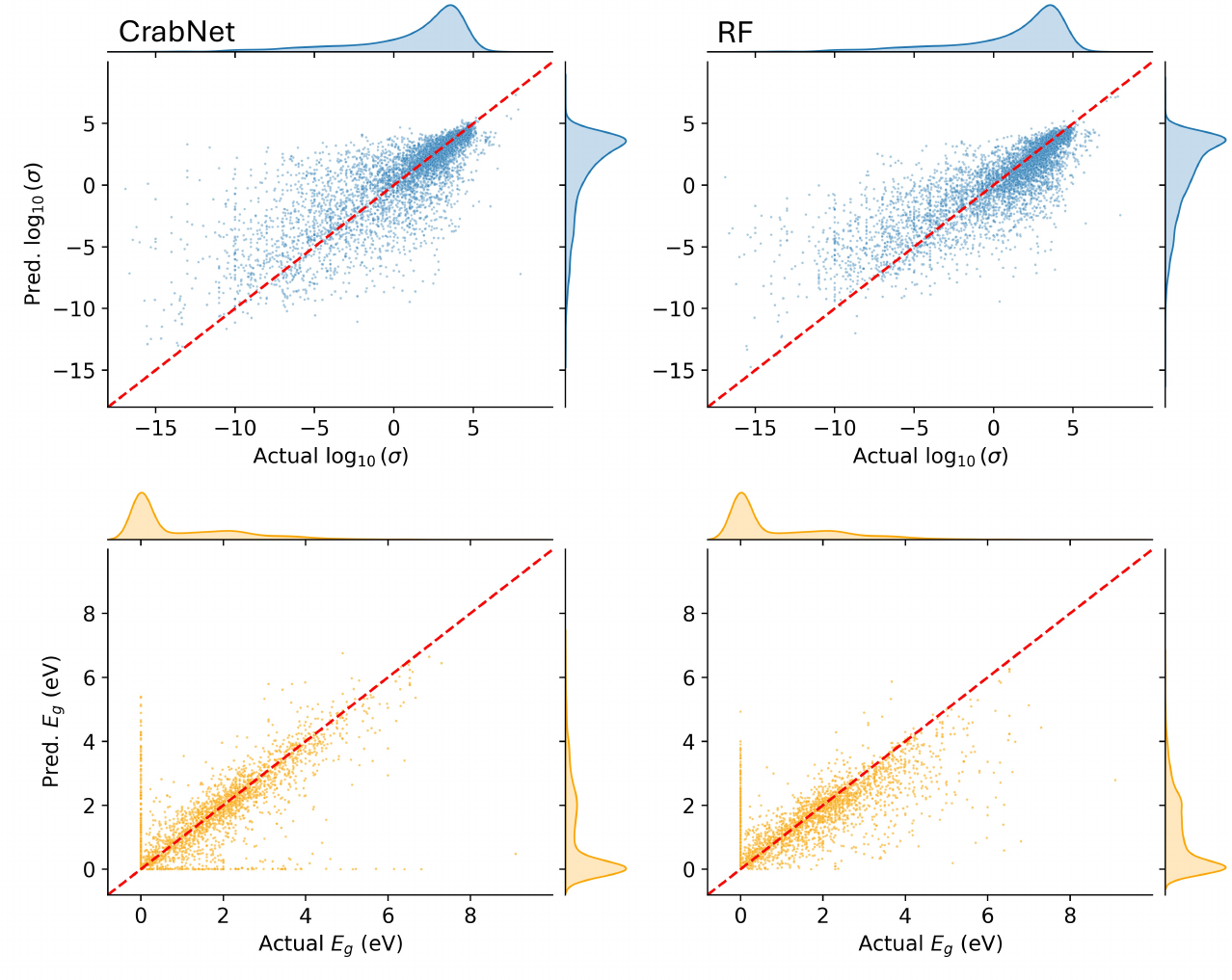}
	\caption{Parity plots are shown for both electrical conductivity (top) and band gap (bottom) prediction. These were obtained by concatenating the different validation folds used in the K-fold evaluation scheme.}
	\label{fig:parity}
\end{figure*}

In Tables \ref{table:sigma} and \ref{table:gap} we report evaluation results for ML prediction on both the properties considered. Figure \ref{fig:loco} illustrates the distinct material clusters obtained for the LOCO-CV evaluation setting. In Figure \ref{fig:parity} , we show parity plots related to the K-fold evaluation scheme.

\paragraph{Conductivity prediction}For electrical conductivity, CrabNet and RF yield comparable in-sample results (K-fold), with RF achieving a $\sim 6\%$ higher $R^2$ than CrabNet, and a slightly improvement over MAE, although not statistically significant. Such an outcome is expected, considering the remarkable performance of RF in interpolation tasks (in-sample). This is due to the intrinsic ensemble nature of the algorithm, enabling a good generalization within the range of training data.
In the out-of-sample evaluation (LOCO-CV), we observe that differences among models are not statistically significant and are subject to high variability. This primarily stems from the size disparities among various material clusters. Additionally, it is plausible that certain material groups contain crucial chemical information that is missing from the training data. The systematic exclusion of such clusters at training stage may lead to a significant degradation in predictive performance, and contribute to an increased variance in the final evaluation.  For example, \textit{cluster 1} depicted in orange in Figure \ref{fig:loco} contains around 95\% of the oxides in the entire dataset. This highlights a scenario where the extrapolation task becomes too demanding for the model, as it is required to identify a great variability across multiple orders of magnitude, all without prior exposure to such conditions in the training dataset.

\paragraph{Band gap prediction} In the case of band gap, it is possible to observe a remarkable improvement of CrabNet compared to RF, with a decrease in MAE of $\sim 27\%$, and a slight average improvement in terms of $R^2$, although not statistically significant. In this scenario, we posit that the adoption of transfer learning provides a significant contribution (see Section \ref{sec:metals_nonmetals} ).
This trend is also partially evident in the LOCO-CV task; nevertheless, once again, the high variability poses challenges for a precise analysis in the out-of-distribution scenario. We believe that increasing the number of clusters can mitigate this issue, by ensuring a more consistent size of the training dataset in each iteration. However, a larger number of clusters increases the likelihood of similar data points being shared between the training and testing datasets, limiting the out-of-distribution assessment. Further exploration of this trade-off will be addressed in future research. 
\subsection{Identification of metals and non-metals}
\label{sec:metals_nonmetals}
Accurate band gap prediction is critical for our ML pipeline. However, challenges arise due to the imbalance between metals and non-metals in material datasets, leading to frequent misclassification of metals as semiconductors or insulators, which can undermine prediction reliability~\citep{riebesell2024pushing}. Various strategies have been explored to mitigate this issue. A first attempt might be partitioning the task into two stages. The initial stage entails training a classifier to discriminate materials into metals and non-metals, eventually using loss-weighting schemes to limit the impact of class imbalances. The next stage would involve a regression task on the subset of non-metals by the preceding classification step. These methods have shown a limited effectiveness in practice \citep{riebesell2024pushing}.
We believe that an interesting alternative may involve foundation models pre-trained on large multi-domain datasets~\citep{shoghi2023molecules}, to be then fine-tuned for specific tasks with limited data~\citep{jablonka2024}. However, we argue that a key concern with foundation models is potential data leakage during pre-training, which can lead to overly optimistic results in downstream tasks.

In our study, to enhance the accuracy of band gap identification, and thus minimizing the number of false negatives (in our definition, \textit{metals} that are wrongly predicted as \textit{semiconductors} or \textit{insulators}), we have utilized a transfer learning approach.
This involved pre-training CrabNet on an extensive dataset sourced from the Materials Project \citep{matproj}, encompassing all entries with chemical formulas and associated band gap information. At the time of writing, 153,224 material entries along with their respective band gaps are present in the Materials Project database. From this initial dataset, we filtered out chemical formulas that were deemed equivalent in our experimental band gap dataset, encompassing 4,767 material entries. We have used the reduced chemical formula as criterion to establish equivalent entries, as atomic proportions are utilized when creating inputs to ML models.
To ensure a fair evaluation we have discarded all such entries, ending up with a pretraining dataset consisting of 149,714 data points. Further processing is conducted on the resulting data to handle duplicates. We utilize a similar strategy akin to that employed for the experimental $E_g$ dataset. Once duplicated material groups are identified, we eliminate those with a standard deviation exceeding 0.1 eV in terms of the corresponding band gaps.
We have used this pool of data to pretrain CrabNet on DFT-calculated band gaps. This is later fine-tuned on our experimental $E_g$ dataset.
\begin{figure*}[!t]
	\centering
	\includegraphics[width=0.98\textwidth]{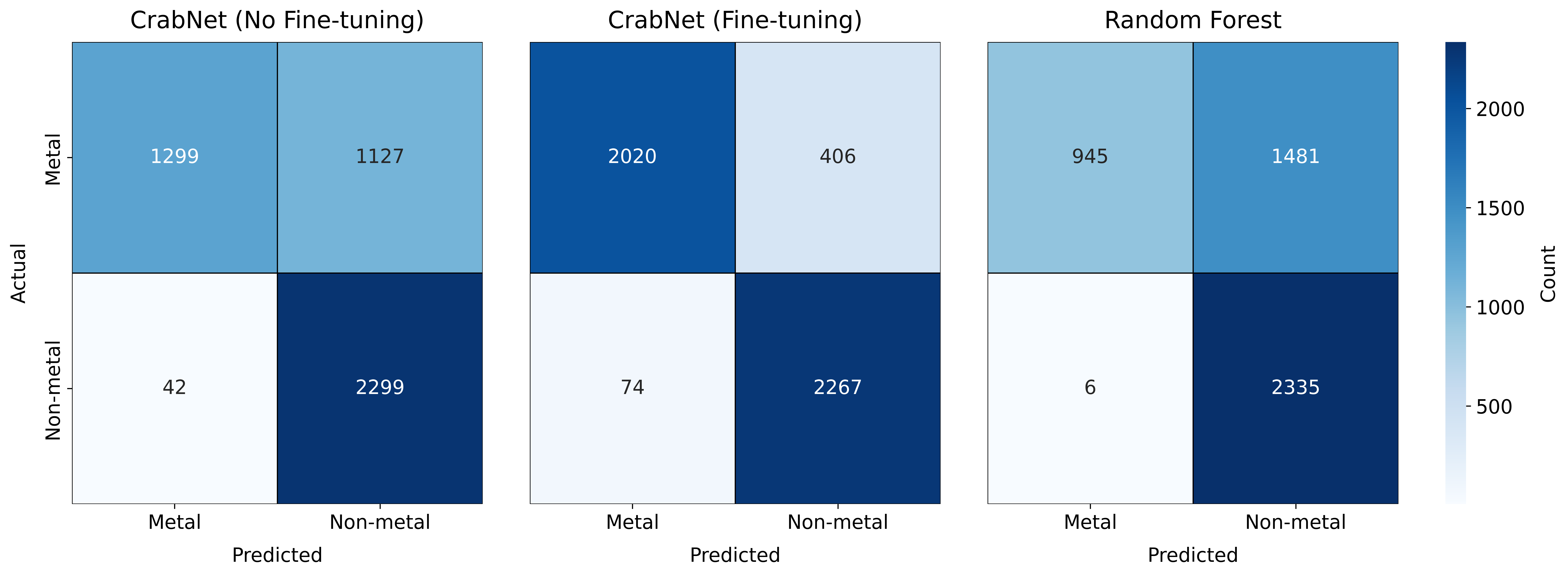}
	\caption{Confusion matrices for the metal vs. non-metal classification task are displayed for the standard CrabNet (left), fine-tuned CrabNet (center), and RF. The fine-tuned CrabNet shows a remarkable improvement, with a significant reduction in false negatives compared to both the standard CrabNet and RF models.}
	\label{fig:ft_gaps}
\end{figure*}
In terms of regression metrics, the fine-tuned model demonstrates enhancements of roughly $\approx 20\%$ in MAE and $ \approx 10\%$ in terms of $R^2$. To better evaluate the fine-tuned model’s effectiveness in reducing false negatives, we investigate the predictions from both the original and fine-tuned models from a classification perspective. For a comprehensive assessment, we also include RF predictions within this evaluation. First, a simple rounding scheme is applied to all the obtained predictions. Specifically, predicted band gaps that are zero when rounded to two decimal places (i.e., values less than $0.005$) are assigned a label of \textit{0}, indicating metals. Predicted band gaps that round to non-zero values (i.e., $0.005$ or greater) are assigned a label of \textit{1}, indicating non-metals.
\begin{figure*}[!t]
	\centering
	\includegraphics[width=0.85\textwidth]{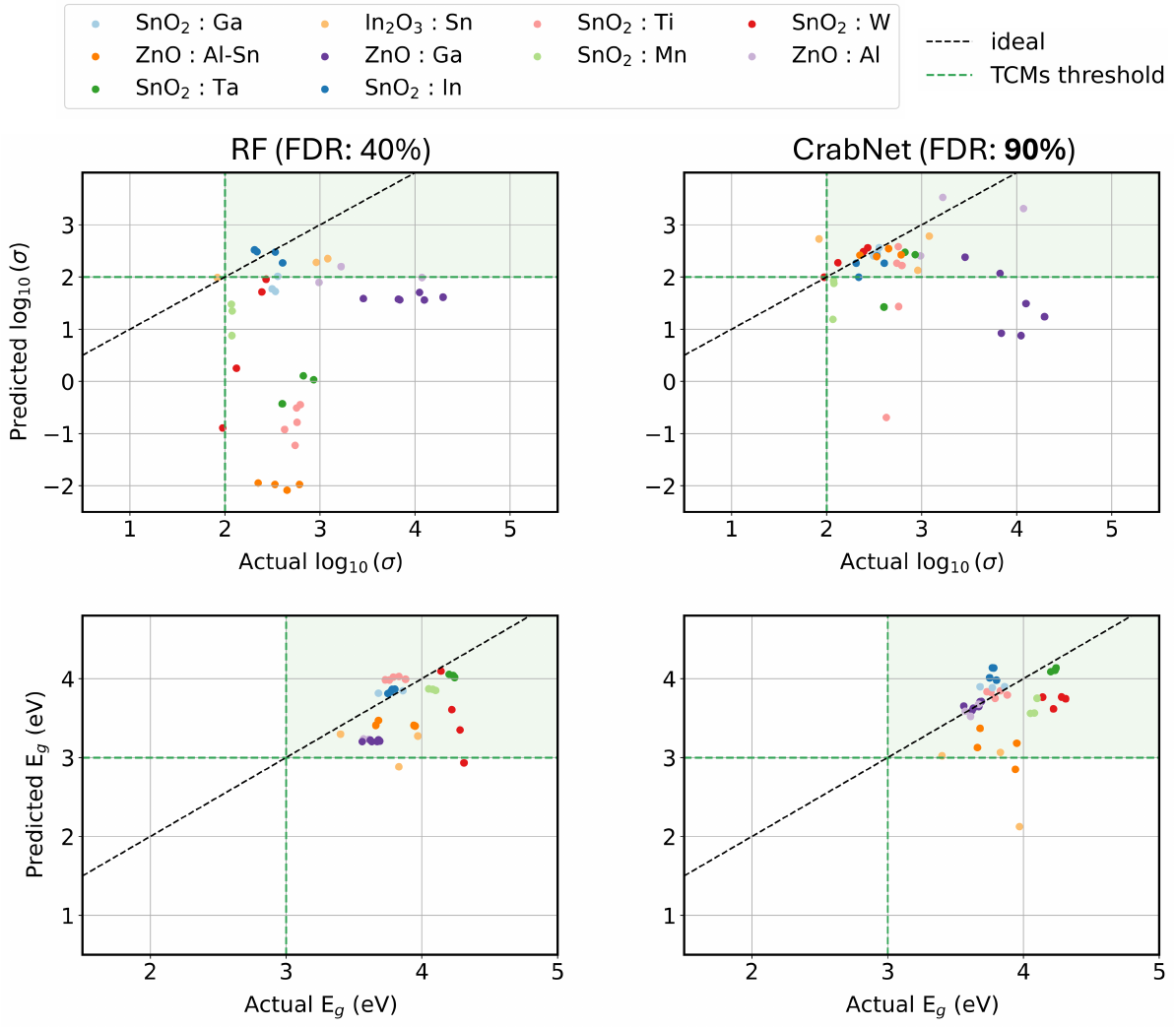}
	\caption{Predicted test TCMs within the leave-one-TCM-family-out evaluation setting, categorized by the constituent properties of electrical conductivity (top) and band gap (bottom). The FDR score indicates the percentage of test TCM families correctly identified by the models, i.e. test materials correctly predicted with respect to the thresholds of $10^2$ S/cm for conductivity, and $3$ eV for band gap.}
	\label{fig:lotcmo_plot}
\end{figure*}
In Figure \ref{fig:ft_gaps} we report the confusion matrices related to the different models considered. In terms of CrabNet, a significant decrease is observed in the count of false negatives, from the initial model (1127) to the fine-tuned one (406). This improvement comes with a slight increase in false positives (instances where semiconductors or insulators are incorrectly predicted as metals), rising from 42 in the model without fine-tuning to 74 in the fine-tuned model.
For RF, we note a significant tendency to overestimate band gaps, resulting in a large number of metals being incorrectly predicted as non-metals (1481). Interestingly, in terms of false positives, only 6 non-metals are misclassified as metals. Further investigation on this aspect is deferred to future research.
Additionally, we utilize \textit{Matthews correlation coefficient} (MCC) \citep{chicco2020mcc} as a robust metric to quantify models' performance on binary classification, given its suitability for imbalanced data. It is defined as follows:
\begin{equation}
	MCC := \frac{(TP \times TN)-(FP \times FN)}{\sqrt{(TP + FP)(TP + FN)(TN + FP)(TN + FN)}} \, ,
\end{equation}
with $TP \, , TN \, , FP \, , FN$ denoting, as usual, \textit{true positives}, \textit{true negatives}, \textit{false positives}, and \textit{false negatives}, respectively. 
A significant improvement is observed when comparing CrabNet without fine-tuning to the fine-tuned version, with the MCC increasing from 0.58 to 0.80. Conversely, the MCC obtained from the RF model is 0.48, which is significantly lower. This can be attributed to the tendency of the model in overestimating the band gaps, leading to a high number of false negatives. Considering the pivotal role that band gap prediction plays in the primary objective of this work, namely accelerating the identification of new TCMs, we believe that this analysis holds fundamental significance. In this context, improving the precision of ML models in discriminating metals from non-metals greatly facilitates the selection of promising material subsets for further investigation.
\subsection{Leave-one-TCM-family-out}
We have discussed the results of two classic evaluation schemes, which carry intrinsic limitations. On the one hand, K-fold provides limited insights on the real possibilities of identifying materials outside the training distribution, frequently yielding overestimated results. On the other hand, LOCO-CV often leads to a noisy evaluation, due to the different size of the obtained material clusters.
In Figure \ref{fig:lotcmo_plot} , we present the results obtained from the proposed leave-one-TCM-family-out benchmark, showcasing the joint predictions of both RF and CrabNet and for both properties under consideration ($\sigma$ and E$_g$).
We immediately notice that CrabNet is the only model capable of identifying the majority of TCMs families in the test set, achieving an FDR of 90\%, compared to 40\% obtained by RF. 
The main challenge results in the identification of electrical conductivity in these materials. As shown in Figure \ref{fig:lotcmo_plot} , RF significantly underestimates this property.
However, in the case of band gap prediction, both models correctly identify over 90\% of the total materials. We believe this is primarily due to the smoother relationship between stoichiometry and band gap, which simplifies the out-of-distribution evaluation. Overall, our analysis shows a superior robustness of CrabNet in identifying novel stoichiometric combinations that were not present in the training distribution.
\begin{figure*}[!t]
	\centering
	\includegraphics[width=0.98\textwidth]{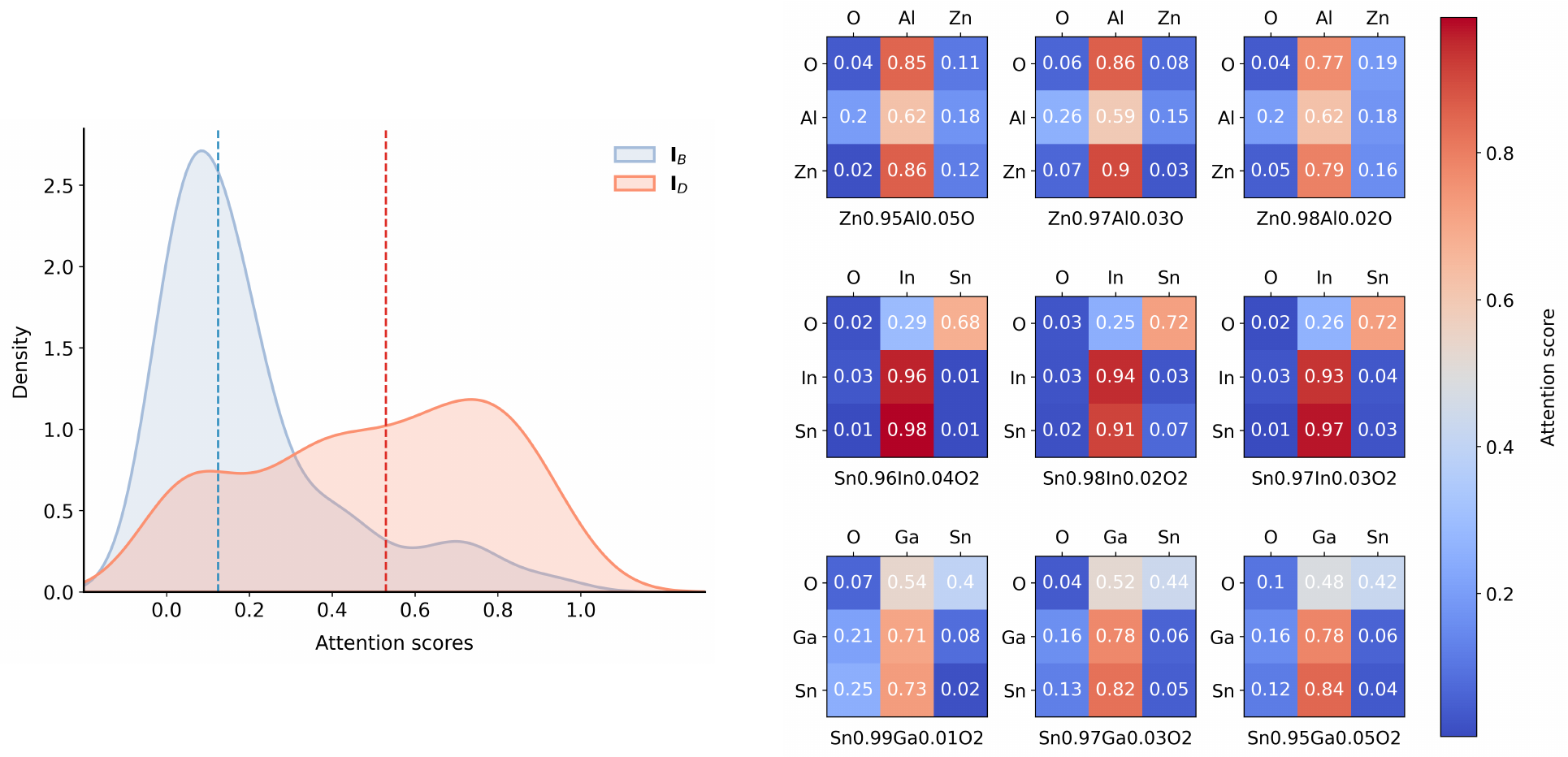}
	\caption{Distributions of attention scores categorized in terms of interaction with base elements $\bm{I}_B$ and with dopants $\bm{I}_D$ (left). Examples of attention matrices extracted for test TCMs in the leave-one-TCM-family out evaluation scheme (right).}
	\label{fig:att_analysis}
\end{figure*}
\subsection{Predictions explainability via attention scores}
Although the significant breakthroughs enabled by deep learning in materials informatics, the \textit{interpretability} of these methods still remains severely limited, giving rise to entire branches of research which aim to improve human understanding of ML models 
(\textit{explainable AI}) \citep{xai2021linardatos}. The interpretability of ML is indeed a crucial aspect, that acquires further importance in scientific applications, often characterized by collaboration among researchers from various fields, and with different backgrounds. However, current approaches often rely on black-box functions, which offer limited insights into the decision-making process.
Notably, the transformer architecture introduced by \citet{vaswani} provides an inherent mechanism for interpreting its decision-making process through the use of self-attention.
The analysis of the underlying attention scores can indeed offer insights about tokens' significance with respect to the surrounding context.

To investigate the superior predictive accuracy achieved by CrabNet in conductivity prediction, we examined the corresponding attention scores generated during the leave-one-TCM-family-out evaluation scheme. Specifically, we extracted attention scores from the last layer of \modif{the} CrabNet encoder, averaged by the corresponding number of attention heads. In this context, we aim to understand whether the model captures complex chemical phenomena related to doping. In this context, we indicate with $B=\{b_1, \dots \, , b_n\}$ the \textit{base} elements, i.e. those which are present in the pristine form of the material, while with $D=\{d_1, \dots \, d_k\}$ we indicate the dopant elements in the chemical formula. For example, for $\text{Zn}_{0.95}\text{Al}_{0.05}\text{O}$ we have $B =\{\text{Zn, O}\}$ and $D = \{\text{Al}\}$ while for $\text{Zn}_{0.97}\text{Al}_{0.02}\text{Sn}_{0.01}\text{O}_2$ we have $B = \{\text{Zn, O}\}$ and $D=\{\text{Al, Sn}\}$. We categorize entries of the attention matrices into four interaction groups:
\begin{itemize}
    \item $\bm{A}_{BB} = [\bm{A}_{ij}]$ with $e_i \, , e_j \in B$ for \textit{base-base} interactions;

    \item $\bm{A}_{BD} = [\bm{A}_{ij}]$ with $e_i \in B \, , d_j \in D$ for \textit{base-dopant} interactions;

    \item $\bm{A}_{DD} = [\bm{A}_{ij}]$ with $d_i \, , d_j \in D$ for \textit{dopant-dopant} interactions;

    \item $\bm{A}_{DB} = [\bm{A}_{ij}]$ with $d_i \in D \, , e_j \in B$ for \textit{dopant-base} interactions.
\end{itemize}
    The interactions involving base elements, $\mathbf{I}_B := \mathbf{A}_{DB} \cup \mathbf{A}_{BB}$, and those involving dopants, $\mathbf{I}_D := \mathbf{A}_{BD} \cup \mathbf{A}_{DD}$, reveal distinct patterns in the attention scores. As shown in Figure~\ref{fig:att_analysis} (left), the distribution of $\mathbf{I}_D$ exhibits a clear shift towards higher attention scores compared to $\mathbf{I}_B$, with the medians indicated by dotted lines. This suggests that the model assigns a greater importance to the interactions involving dopants, effectively capturing their critical role in shaping material representations for conductivity prediction.
\section{Testing the search for new TCMs}
To assess ML models’ effectiveness in identifying TCMs, a search was conducted in the Pearson’s Crystallographic Database~\citep{pearson}, MPDS~\citep{mpds} and ICSD~\citep{icsd} for compounds containing elements commonly found in known classes of TCMs. Predicting their properties with ML could reveal materials previously overlooked as TCMs. For this experiment, we utilize CrabNet, given its good performance in the proposed leave-one-TCM-family-out evaluation method.
We conducted a search for oxide compounds containing combinations of three cations from Zn, Ga, Sn, Al, and In.
We also include a small selection of five compositions across MPDS and ICSD of doped binary oxides (ZnO, $\text{SnO}_2$ and $\text{In}_2\text{O}_3$), with dopants not present in the training dataset.
We end up with a final list comprising 55 compositions shown in Table \ref{fig:screen_table}.  

We utilize the same TCMs criteria established for the the leave-one-TCM-family-out evaluation. Specifically, we are targeting materials with band gap $E_g > 3$ eV and with conductivity $\sigma > 10^2$ S/cm. Compositions meeting these criteria are highlighted in Table \ref{fig:screen_table}.
To provide a global assessment of ML-predicted materials, we define a figure of merit $\Phi_M$ as:
\begin{equation}
	\Phi_M = \hat{E}_g \cdot \hat{\sigma} \, ,
\end{equation}
where $\hat{E}_g$ and $\hat{\sigma}$ denote the predicted band gap and conductivity (as $\log_{10}$) from ML models' ensembles, respectively. In essence, $\Phi_M$ will prioritize an optimal trade-off between the two properties.
We further utilize a risk-adjusted figure of merit $\Phi^{std\text{-}adj}_M$ \citep{riebesell2024pushing}, defined as
\begin{equation}
\Phi^{std\text{-}adj}_M := \Phi_M - \Phi^{std}_M \, ,
\end{equation}
where $\Phi^{std}_M = \sqrt{\hat{E}^2_g \, s^2_{\hat{\sigma}} + \hat{\sigma}^2 \, s^2_{\hat{E}_g}}$ is obtained by uncertainty propagation rules for multiplication, and $s_P$ denotes the uncertainty produced by an ensemble of ML models (standard deviation corresponding to the predictive mean, see Appendix \ref{app:unc_est}) for a predicted property $P$. The risk-adjusted figure of merit $\Phi^{std\text{-}adj}_M$ is essentially defined by subtracting one standard deviation from the original figure of merit $\Phi_M$. Compositions with high figure of merit and low uncertainty in their prediction are prioritised over compositions with large uncertainty in their prediction.

From the analysis of the model outputs of the 55 materials selected above, the compositions with the highest  $\Phi^{std\text{-}adj}_M$ are, as expected, those most similar to the training dataset. Doped binary oxides are ranked high by $\Phi_M$ and their band gaps are accurately predicted. $\text{Na}_{0.025}\text{Zn}_{0.975}\text{O}_{0.988}$ (entry \textit{1}) is predicted to have a conductivity of 3.57 $\log_{10}(\sigma)$ S/cm, although measurements reported in the literature are much lower, due to the low concentration of p-type carriers~\citep{erdogan2021effect}. The band gap prediction of $\text{Na}_{0.025}\text{Zn}_{0.975}\text{O}_{0.988}$ (3.74 eV), compares to the reported experimental measurement of $3.26$ eV~\citep{basyooni2017enhanced}. Thin films of $\text{Ca}_{0.04}\text{Zn}_{0.96}\text{O}$ (entry \textit{2}) exhibit a band gap of 3.40 eV and a conductivity of $1.3 \, \log_{10}(\sigma)$ S/cm~\citep{mahdhi2018synthesis}. CrabNet predicts a band gap of 4.02 eV and a conductivity of $3.22 \, \log_{10}(\sigma)$ S/cm for this composition.
\modif{Both the band gap and conductivity predictions show consistency with the expected error ranges outlined in Table \ref{table:sigma} and Table \ref{table:gap}. The higher deviation in conductivity predictability is considered acceptable given the inherent complexity of predicting conductivity solely from stoichiometry.} Notably, neither $\{\text{Ca, Zn, O}\}$ nor $\{\text{Na, Zn, O}\}$ phase fields are present in the training dataset. For materials containing three cations, indium-containing phase fields rank near the top (Table~\ref{fig:screen_table}). This is expected, given the well-established significance of $\text{In}_2\text{O}_3$ in the TCMs literature.
Materials in the $\text{Ga}_{2}\text{O}_{3}$-$\text{In}_{2}\text{O}_{3}$-$\text{SnO}_2$ phase field~\citep{edwards1998subsolidus} have been explored as transparent conductors~\citep{edwards1997new}, with the highest-ranking material in the phase field $\text{Ga}_{0.06}\text{In}_{1.92}\text{Sn}_{0.02}\text{O}_{3.01}$ (entry \textit{7}) having a reported conductivity of $3.43 \, \log_{10}(\sigma)$ S/cm and a band gap of $3.04$ eV~\citep{dolgonos2015phase} which the model does well at predicting with $2.92$ $\log_{10}(\sigma)$ S/cm for conductivity and $3.52$ eV for band gap.
Among the highest $\Phi^{std\text{-}adj}_M$-ranked materials, $\text{Al}_{0.67}\text{Ga}_{1.33}\text{Zn}_{37}\text{O}_{40}$ (entry \textit{3}) is a homologous phase $\text{((Ga}_{1-\alpha}\text{Al}_\alpha\text{)}_2\text{O}_3\text{(ZnO)}_m\text{)}$ in the pseudo-ternary $\text{Ga}_{2}\text{O}_{3}\text{-Al}_{2}\text{O}_{3}\text{-ZnO}$ phase field and has been postulated as a potential thermoelectric~\citep{michiue2020utilizing} but not as a TCM, and its band gap and electrical conductivity were not reported. Given that other materials in the $\text{Ga}_{2}\text{O}_{3}\text{-Al}_{2}\text{O}_{3}\text{-ZnO}$ phase field have very high conductivity ($1.0 \times 10^4$ to $1.6 \times 10^4$ S/cm)~\citep{michiue2020utilizing} it could be expected that the composition $\text{Al}_{0.67}\text{Ga}_{1.33}\text{Zn}_{37}\text{O}_{40}$ could also show high conductivity and an appropriate band gap. The Al doped $\text{Zn}_{2}\text{SnO}_{4}$ spinel, $\text{Al}_{0.04}\text{Sn}_{0.98}\text{Zn}_{1.98}\text{O}_{4}$ (entry \textit{21} in Table \ref{fig:screen_table}) has been explored as TCM for CIGS solar cells~\citep{jung2020doped}, and has a measured band gap of $> 3.5$ eV but low conductivity ($1.11 \log_{10}(\sigma)$ S/cm). Other spinel materials have had their conductivity measured, for example $\text{GaInZnO}_{4}$ (entry \textit{51}) has a measured conductivity of 2.7 $\log_{10}(\sigma)$ S/cm which is much higher than predicted (-6.6 $\log_{10}(\sigma)$ S/cm), and a band gap of $3.5$ eV~\citep{orita2000mechanism} which is predicted very closely ($3.26$ eV).  
\FloatBarrier
\begin{table*}[!t]
	\centering
	\includegraphics[width=0.75\textwidth]{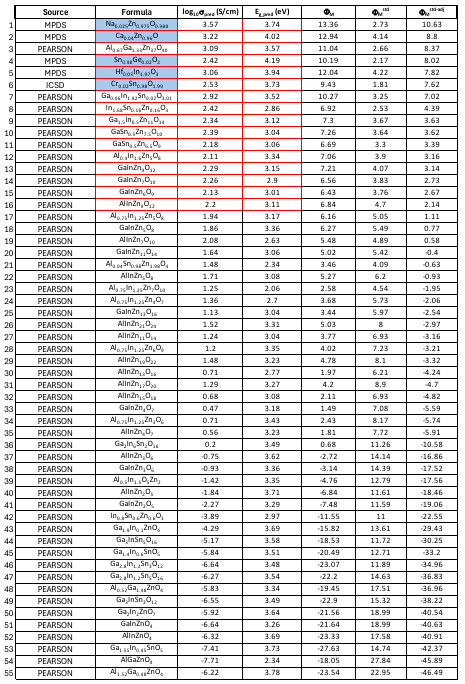}
    \caption{Predicted values of conductivity ($\sigma$) and band gap ($E_g$) for a set of materials containing elements common to known classes of TCMs i.e. oxides with combinations of Zn, Al, Ga, In and Sn, with additional dopant elements. $\Phi_M$, $\Phi^{std}_M$ and $\Phi^{std\text{-}adj}_M$ are figure of merit values as discussed in the main text. Databases in which the compositions were searched are given in the \textit{Source} column. Red-bordered cells indicate materials meeting our TCMs criteria, with a conductivity greater than $2 \, \log_{10}(\sigma)$ S/cm, and a band gap greater than $3$ eV. Rows with formula high-lighted in blue are doped binary oxides of Zn, Sn or In, closest to the training dataset; unhighlighted rows are three cation oxide materials with constituent elements commonly found within well-known TCM classes. Compositions are ordered from highest to lowest $\Phi^{std\text{-}adj}_M$}
    \label{fig:screen_table}
\end{table*}
\FloatBarrier
These two examples show that the model recognizes that doping small amounts of elements into structures can induce conductivity ($\text{Al}_{0.04}\text{Sn}_{0.98}\text{Zn}_{1.98}\text{O}_{4}$) and more stoichiometric closed shell materials are less likely to display conductivity ($\text{GaInZnO}_{4}$). In fact, the measured conductivity in $\text{GaInZnO}_{4}$ results from $\text{Ga}$ anti-site defects, $\text{Ga}_{\text{Zn}}$, as the major electron donor in $\text{GaInZnO}_{4}$~\citep{altynbek2013carrier} which would be difficult for an ML model to capture, when trained on composition only.
\modif{This is because the oxidation states present would correspond to filled bands and thus to a low conductivity in terms of electron count, while the model is unable to recognise the self-doping that produces the experimentally observed conductivity.}

\section{Limitations}
While we believe our analysis has provided valuable insights into leveraging data-driven methods for accelerating the discovery of new TCMs, it is important to acknowledge certain limitations inherent in our approach. 

The first challenge stems from the inherently limited pool of existing TCMs. Given the scarcity of such materials in current databases or literature, our data-driven pipeline is inevitably constrained, impacting the breadth and depth of the proposed analysis. 

A second limitation relates to the specific mechanisms underlying the properties of the materials of interest. If the goal is to identify TCMs similar to those in the training dataset, 
the proposed framework is indeed promising, as shown in Table~\ref{fig:screen_table}. However, when seeking materials that achieve the desired properties through different mechanisms, our approach is less likely to provide new insights into the underlying Chemistry. This is because data-driven frameworks largely depends on the patterns reflected in the training dataset, which may not capture the diversity of mechanisms outside the established categories. This limitation was already highlighted in the work of \citet{ml_extrap_mats} and \citet{schrier_exceptional}.

Another limitation arises from the nature of the input data used in this study, which focuses solely on the stoichiometry of materials.
In exploratory settings, stoichiometry-based methods provide a valuable and natural baseline since structural information is typically unavailable. However, when additional information is available, it becomes essential to incorporate it effectively. Moving forward, we foresee the integration of more detailed prior knowledge as an important next step. This could involve leveraging recent developments in Large Language Models (LLMs) to encode domain knowledge in chemistry, as suggested by \cite{xie2023large} and \cite{jablonka2024}, or incorporating structural data via representation learning schemes \citep{lee2023clcs, lee2023stoichiometry}.
\section{Conclusions}
We have proposed a bespoke data-driven framework aimed at leveraging data-driven methods to accelerate the discovery of new TCMs. To address the challenge of limited and sparse material data, we created two experimental datasets of room-temperature conductivity and band gap.
This involved the collection of raw data, followed by the application of a meticulous, line-by-line validation to verify the correctness of the reported chemical formulas, alongside the corresponding measurements of electrical conductivity and band gap.
The validated datasets were used as foundation for evaluating SOTA ML models for property-prediction from the stoichiometry alone. We have proposed a bespoke evaluation method to empirically measure the potential of ML in identifying new classes of TCMs.
Finally, we have compiled a list of 55 compositions sourced across various material databases, to test the effectiveness of ML in accelerate the identification of new TCMs.
Overall, our results suggest that ML has the potential to identify new TCMs that are compositionally similar to the ones in the training dataset. Nonetheless, we argue that this holds significant value, as it enables an accelerated identification of compounds that may have been previously overlooked as TCMs.
\section{Implementation details}\label{app:impl}
CrabNet has been implemented with a batch size of 512, a RobustL1 loss function, a Lamb Lookahead optimizer with stochastic weight averaging, a cyclic learning rate from $1 \times 10^{-4}$ to $6 \times 10^{-3}$. 
For RF, we have utilized a modified \verb|sci-kit learn| implementation, which produces aleatoric and epistemic contributions to uncertainty \citep{riebesell_thermoelectric}. Apart from that, we have utilized default factory settings as in the original \verb|RandomForestRegressor| class in the \verb|sci-kit learn| implementation.

\section{Data availability statement}
\modif{The original, unmodified band gap dataset, which serves as the basis for the enriched version proposed in this study, is available in the publication by \citet{zhuo_2018}. }
The electrical conductivity dataset was compiled from raw data obtained from the UCSB repository and the Materials Platform for Data Science (MPDS). The UCSB repository can be accessed via the following link: \href{https://hackingmaterials.lbl.gov/matminer/dataset_summary.html}{https://hackingmaterials.lbl.gov/matminer/dataset\_summary.html} (\verb|ucsb_thermoelectrics|), while an API license for MPDS can be purchased at \href{https://mpds.io/}{https://mpds.io/} .

\section{Acknowledgements}
\modif{We thank EPSRC for support under EP/V026887/1 and the Impact Acceleration Account, Pilkington (NSG Group) and Leverhulme Trust through the Leverhulme Research Centre for Functional Materials Design (RC-2015-036).}

\bibliographystyle{unsrtnat}
\bibliography{references}  

\begin{thebibliography}{95}
\providecommand{\natexlab}[1]{#1}
\providecommand{\url}[1]{\texttt{#1}}
\expandafter\ifx\csname urlstyle\endcsname\relax
  \providecommand{\doi}[1]{doi: #1}\else
  \providecommand{\doi}{doi: \begingroup \urlstyle{rm}\Url}\fi

\bibitem[Schleder et~al.(2019)Schleder, Padilha, Acosta, Costa, and
  Fazzio]{from_dft_to_ml}
Gabriel~R Schleder, Antonio C~M Padilha, Carlos~Mera Acosta, Marcio Costa, and
  Adalberto Fazzio.
\newblock From $\text{DFT}$ to machine learning: recent approaches to materials
  science–a review.
\newblock \emph{Journal of Physics: Materials}, 2\penalty0 (3):\penalty0
  032001, may 2019.
\newblock \doi{10.1088/2515-7639/ab084b}.
\newblock URL \url{https://dx.doi.org/10.1088/2515-7639/ab084b}.

\bibitem[Pollice et~al.(2021)Pollice, dos Passos~Gomes, Aldeghi, Hickman,
  Krenn, Lavigne, Lindner-D'Addario, Nigam, Ser, Yao, and
  Aspuru-Guzik]{data_driven_mat_design}
Robert Pollice, Gabriel dos Passos~Gomes, Matteo Aldeghi, Riley~J. Hickman,
  Mario Krenn, Cyrille Lavigne, Michael Lindner-D'Addario, AkshatKumar Nigam,
  Cher~Tian Ser, Zhenpeng Yao, and Al{\'a}n Aspuru-Guzik.
\newblock Data-driven strategies for accelerated materials design.
\newblock \emph{Accounts of Chemical Research}, 54\penalty0 (4):\penalty0
  849--860, Feb 2021.
\newblock ISSN 0001-4842.
\newblock \doi{10.1021/acs.accounts.0c00785}.
\newblock URL \url{https://doi.org/10.1021/acs.accounts.0c00785}.

\bibitem[Jain et~al.(2013)Jain, Ong, Hautier, Chen, Richards, Dacek, Cholia,
  Gunter, Skinner, Ceder, and Persson]{matproj}
Anubhav Jain, Shyue~Ping Ong, Geoffroy Hautier, Wei Chen, William~Davidson
  Richards, Stephen Dacek, Shreyas Cholia, Dan Gunter, David Skinner, Gerbrand
  Ceder, and Kristin~a. Persson.
\newblock {The Materials Project: A materials genome approach to accelerating
  materials innovation}.
\newblock \emph{APL Materials}, 1\penalty0 (1):\penalty0 011002, 2013.
\newblock ISSN 2166532X.
\newblock \doi{10.1063/1.4812323}.
\newblock URL \url{http://link.aip.org/link/AMPADS/v1/i1/p011002/s1\&Agg=doi}.

\bibitem[Blokhin and Villars(2018)]{mpds}
Evgeny Blokhin and Pierre Villars.
\newblock \emph{The PAULING FILE Project and Materials Platform for Data
  Science: From Big Data Toward Materials Genome}, pages 1--26.
\newblock Springer International Publishing, Cham, 2018.
\newblock ISBN 978-3-319-42913-7.
\newblock \doi{10.1007/978-3-319-42913-7_62-1}.
\newblock URL \url{https://doi.org/10.1007/978-3-319-42913-7_62-1}.

\bibitem[Kirklin et~al.(2015)Kirklin, Saal, Meredig, Thompson, Doak, Aykol,
  R{\"u}hl, and Wolverton]{OQMD}
Scott Kirklin, James~E Saal, Bryce Meredig, Alex Thompson, Jeff~W Doak,
  Muratahan Aykol, Stephan R{\"u}hl, and Chris Wolverton.
\newblock The open quantum materials database (oqmd): assessing the accuracy of
  $\text{DFT}$ formation energies.
\newblock \emph{npj Computational Materials}, 1\penalty0 (1):\penalty0 15010,
  2015.
\newblock \doi{10.1038/npjcompumats.2015.10}.
\newblock URL \url{https://doi.org/10.1038/npjcompumats.2015.10}.

\bibitem[Riebesell et~al.(2024)Riebesell, Surta, Goodall, Gaultois,
  et~al.]{riebesell2024pushing}
Janosh Riebesell, Todd~Wesley Surta, Rhys Edward~Andrew Goodall,
  Michael~William Gaultois, et~al.
\newblock Discovery of high-performance dielectric materials with
  machine-learning-guided search.
\newblock \emph{Cell Reports Physical Science}, 2024.

\bibitem[Conduit et~al.(2019)Conduit, Illston, Baker, Duggappa, Harding, Stone,
  and Conduit]{ml_nickel_superallo}
B.D. Conduit, T.~Illston, S.~Baker, D.~Vadegadde Duggappa, S.~Harding, H.J.
  Stone, and G.J. Conduit.
\newblock Probabilistic neural network identification of an alloy for direct
  laser deposition.
\newblock \emph{Materials \& Design}, 168:\penalty0 107644, 2019.
\newblock ISSN 0264-1275.
\newblock \doi{https://doi.org/10.1016/j.matdes.2019.107644}.
\newblock URL
  \url{https://www.sciencedirect.com/science/article/pii/S0264127519300814}.

\bibitem[Mansouri~Tehrani et~al.(2018)Mansouri~Tehrani, Oliynyk, Parry, Rizvi,
  Couper, Lin, Miyagi, Sparks, and Brgoch]{ml_superhard}
Aria Mansouri~Tehrani, Anton~O. Oliynyk, Marcus Parry, Zeshan Rizvi, Samantha
  Couper, Feng Lin, Lowell Miyagi, Taylor~D. Sparks, and Jakoah Brgoch.
\newblock Machine learning directed search for ultraincompressible, superhard
  materials.
\newblock \emph{Journal of the American Chemical Society}, 140\penalty0
  (31):\penalty0 9844--9853, 2018.
\newblock \doi{10.1021/jacs.8b02717}.
\newblock URL \url{https://doi.org/10.1021/jacs.8b02717}.
\newblock PMID: 30010335.

\bibitem[Schrier et~al.(2023)Schrier, Norquist, Buonassisi, and
  Brgoch]{schrier_exceptional}
Joshua Schrier, Alexander~J. Norquist, Tonio Buonassisi, and Jakoah Brgoch.
\newblock In pursuit of the exceptional: Research directions for machine
  learning in chemical and materials science.
\newblock \emph{Journal of the American Chemical Society}, 145\penalty0
  (40):\penalty0 21699--21716, 10 2023.
\newblock \doi{10.1021/jacs.3c04783}.
\newblock URL \url{https://doi.org/10.1021/jacs.3c04783}.

\bibitem[Way et~al.(2019)Way, Luke, Evans, Li, Kim, Durrant, Hin~Lee, and
  Tsoi]{most_used_tco2}
Amirah Way, Joel Luke, Alex~D. Evans, Zhe Li, Ji-Seon Kim, James~R. Durrant,
  Harrison~Ka Hin~Lee, and Wing~C. Tsoi.
\newblock {Fluorine doped tin oxide as an alternative of indium tin oxide for
  bottom electrode of semi-transparent organic photovoltaic devices}.
\newblock \emph{AIP Advances}, 9\penalty0 (8):\penalty0 085220, 08 2019.
\newblock ISSN 2158-3226.
\newblock \doi{10.1063/1.5104333}.
\newblock URL \url{https://doi.org/10.1063/1.5104333}.

\bibitem[Maurya et~al.(2022)Maurya, Galvan, Gautam, and Xu]{recent_tcm_photo}
Sandeep~Kumar Maurya, Hazel~Rose Galvan, Gaurav Gautam, and Xiaojie Xu.
\newblock Recent progress in transparent conductive materials for
  photovoltaics.
\newblock \emph{Energies}, 15\penalty0 (22), 2022.
\newblock ISSN 1996-1073.
\newblock \doi{10.3390/en15228698}.
\newblock URL \url{https://www.mdpi.com/1996-1073/15/22/8698}.

\bibitem[Morales-Masis et~al.(2017)Morales-Masis, De~Wolf, Woods-Robinson,
  Ager, and Ballif]{tcms_optoelectronics}
Monica Morales-Masis, Stefaan De~Wolf, Rachel Woods-Robinson, Joel~W. Ager, and
  Christophe Ballif.
\newblock Transparent electrodes for efficient optoelectronics.
\newblock \emph{Advanced Electronic Materials}, 3\penalty0 (5):\penalty0
  1600529, 2017.
\newblock \doi{https://doi.org/10.1002/aelm.201600529}.
\newblock URL
  \url{https://onlinelibrary.wiley.com/doi/abs/10.1002/aelm.201600529}.

\bibitem[Woods-Robinson et~al.(2018)Woods-Robinson, Broberg, Faghaninia, Jain,
  Dwaraknath, and Persson]{woodsRob}
Rachel Woods-Robinson, Danny Broberg, Alireza Faghaninia, Anubhav Jain,
  Shyam~S. Dwaraknath, and Kristin~A. Persson.
\newblock Assessing high-throughput descriptors for prediction of transparent
  conductors.
\newblock \emph{Chemistry of Materials}, 30\penalty0 (22):\penalty0 8375--8389,
  2018.
\newblock \doi{10.1021/acs.chemmater.8b03529}.
\newblock URL \url{https://doi.org/10.1021/acs.chemmater.8b03529}.

\bibitem[Xiong et~al.(2018)Xiong, Sun, Xing, and Hato]{Xiong}
X.~Xiong, Y.~Sun, Y.~Xing, and T.~Hato.
\newblock {Investigate Machine Learning Methods for Transparent Conductors
  Prediction}.
\newblock 2018.
\newblock URL \url{http://noiselab.ucsd.edu/ECE228_2018/}.

\bibitem[{Sutton} et~al.(2018){Sutton}, {Bartel}, {Liu}, {Boley}, {Rupp},
  {Ghiringhelli}, {Scheffler}, and {Christopher Sutton Team}]{Sutton}
Christopher {Sutton}, Christopher {Bartel}, Xiangyue {Liu}, Mario {Boley},
  Matthias {Rupp}, Luca {Ghiringhelli}, Matthias {Scheffler}, and {Christopher
  Sutton Team}.
\newblock {Evaluation of Machine Learning Methods for the Prediction of Key
  Properties for Novel Transparent Semiconductors}.
\newblock In \emph{APS March Meeting Abstracts}, volume 2018 of \emph{APS
  Meeting Abstracts}, page E34.012, January 2018.

\bibitem[Villars et~al.(2007)Villars, Cenzual, and Pearson]{pearson}
Pierre Villars, Karin Cenzual, and William~B. Pearson.
\newblock Pearson's crystal data : crystal structure database for inorganic
  compounds.
\newblock 2007.

\bibitem[Rühl(2019)]{icsd}
Stephan Rühl.
\newblock \emph{The Inorganic Crystal Structure Database ( ICSD ): A Tool for
  Materials Sciences}, pages 41--54.
\newblock 10 2019.
\newblock ISBN 9783527341214.
\newblock \doi{10.1002/9783527802265.ch2}.

\bibitem[Hautier et~al.(2014)Hautier, Miglio, Waroquiers, Rignanese, and
  Gonze]{hautier_mass}
Geoffroy Hautier, Anna Miglio, David Waroquiers, Gian-Marco Rignanese, and
  Xavier Gonze.
\newblock How does chemistry influence electron effective mass in oxides? a
  high-throughput computational analysis.
\newblock \emph{Chemistry of Materials}, 26\penalty0 (19):\penalty0 5447--5458,
  10 2014.
\newblock \doi{10.1021/cm404079a}.
\newblock URL \url{https://doi.org/10.1021/cm404079a}.

\bibitem[Sun et~al.(2018)Sun, Xing, Xiong, and Hao]{tcms_ml}
Yan Sun, Yiyuan Xing, Xufan Xiong, and Tianduo Hao.
\newblock Investigate machine learning methods for transparent conductors
  prediction, 2018.
\newblock URL \url{http://noiselab.ucsd.edu/ECE228_2018/Reports/Report3.pdf}.

\bibitem[Antunes et~al.(2022)Antunes, Vikram, Plata, Powell, Butler, and
  Grau-Crespo]{ml_thermo}
Luis~M. Antunes, Vikram, Jose~J. Plata, Anthony~V. Powell, Keith~T. Butler, and
  Ricardo Grau-Crespo.
\newblock \emph{Machine Learning Approaches for Accelerating the Discovery of
  Thermoelectric Materials}, volume 1416 of \emph{ACS Symposium Series},
  doi:10.1021/bk-2022-1416.ch001~1, pages 1--32.
\newblock American Chemical Society, 2023/03/22 2022.
\newblock ISBN 9780841297630.
\newblock \doi{doi:10.1021/bk-2022-1416.ch001}.
\newblock URL \url{https://doi.org/10.1021/bk-2022-1416.ch001}.

\bibitem[Gaultois et~al.(2013)Gaultois, Sparks, Borg, Seshadri, Bonificio, and
  Clarke]{ucsb}
Michael~W. Gaultois, Taylor~D. Sparks, Christopher K.~H. Borg, Ram Seshadri,
  William~D. Bonificio, and David~R. Clarke.
\newblock Data-driven review of thermoelectric materials: Performance and
  resource considerations.
\newblock \emph{Chemistry of Materials}, 25\penalty0 (15):\penalty0 2911--2920,
  Aug 2013.
\newblock ISSN 0897-4756.
\newblock \doi{10.1021/cm400893e}.
\newblock URL \url{https://doi.org/10.1021/cm400893e}.

\bibitem[Zhuo et~al.(2018)Zhuo, Mansouri~Tehrani, and Brgoch]{zhuo_2018}
Ya~Zhuo, Aria Mansouri~Tehrani, and Jakoah Brgoch.
\newblock Predicting the band gaps of inorganic solids by machine learning.
\newblock \emph{The Journal of Physical Chemistry Letters}, 9\penalty0
  (7):\penalty0 1668--1673, 2018.
\newblock \doi{10.1021/acs.jpclett.8b00124}.
\newblock URL \url{https://doi.org/10.1021/acs.jpclett.8b00124}.
\newblock PMID: 29532658.

\bibitem[Wang et~al.(2022)Wang, Zhang, Thé, and Yu]{wang_2022}
Teng Wang, Kefei Zhang, Jesse Thé, and Hesheng Yu.
\newblock Accurate prediction of band gap of materials using stacking machine
  learning model.
\newblock \emph{Computational Materials Science}, 201:\penalty0 110899, 2022.
\newblock ISSN 0927-0256.
\newblock \doi{https://doi.org/10.1016/j.commatsci.2021.110899}.
\newblock URL
  \url{https://www.sciencedirect.com/science/article/pii/S0927025621006078}.

\bibitem[Mukherjee et~al.(2020{\natexlab{a}})Mukherjee, Satsangi, and
  Singh]{mukherjee}
Madhubanti Mukherjee, Swanti Satsangi, and Abhishek~K. Singh.
\newblock A statistical approach for the rapid prediction of electron
  relaxation time using elemental representatives, Jul 2020{\natexlab{a}}.
\newblock URL
  \url{https://acs.figshare.com/collections/A_Statistical_Approach_for_the_Rapid_Prediction_of_Electron_Relaxation_Time_Using_Elemental_Representatives/5071658/1}.

\bibitem[Na et~al.(2021)Na, Jang, and Chang]{na2022dop}
Gyoung~S. Na, Seunghun Jang, and Hyunju Chang.
\newblock Predicting thermoelectric properties from chemical formula with
  explicitly identifying dopant effects.
\newblock \emph{npj Computational Materials}, 7\penalty0 (1):\penalty0 106,
  2021.
\newblock \doi{10.1038/s41524-021-00564-y}.
\newblock URL \url{https://doi.org/10.1038/s41524-021-00564-y}.

\bibitem[Ricci et~al.(2017)Ricci, Chen, Aydemir, Snyder, Rignanese, Jain, and
  Hautier]{Ricci}
Francesco Ricci, Wei Chen, Umut Aydemir, G.~Jeffrey Snyder, Gian-Marco
  Rignanese, Anubhav Jain, and Geoffroy Hautier.
\newblock An ab initio electronic transport database for inorganic materials.
\newblock \emph{Scientific Data}, 4\penalty0 (1):\penalty0 170085, 2017.
\newblock \doi{10.1038/sdata.2017.85}.
\newblock URL \url{https://doi.org/10.1038/sdata.2017.85}.

\bibitem[Choudhary et~al.(2020)Choudhary, Garrity, Reid, DeCost, Biacchi,
  Hight~Walker, Trautt, Hattrick-Simpers, Kusne, Centrone, Davydov, Jiang,
  Pachter, Cheon, Reed, Agrawal, Qian, Sharma, Zhuang, Kalinin, Sumpter,
  Pilania, Acar, Mandal, Haule, Vanderbilt, Rabe, and Tavazza]{jarvis_dft}
Kamal Choudhary, Kevin~F. Garrity, Andrew C.~E. Reid, Brian DeCost, Adam~J.
  Biacchi, Angela~R. Hight~Walker, Zachary Trautt, Jason Hattrick-Simpers,
  A.~Gilad Kusne, Andrea Centrone, Albert Davydov, Jie Jiang, Ruth Pachter,
  Gowoon Cheon, Evan Reed, Ankit Agrawal, Xiaofeng Qian, Vinit Sharma, Houlong
  Zhuang, Sergei~V. Kalinin, Bobby~G. Sumpter, Ghanshyam Pilania, Pinar Acar,
  Subhasish Mandal, Kristjan Haule, David Vanderbilt, Karin Rabe, and Francesca
  Tavazza.
\newblock The joint automated repository for various integrated simulations
  (jarvis) for data-driven materials design.
\newblock \emph{npj Computational Materials}, 6\penalty0 (1):\penalty0 173,
  2020.
\newblock \doi{10.1038/s41524-020-00440-1}.
\newblock URL \url{https://doi.org/10.1038/s41524-020-00440-1}.

\bibitem[Yao et~al.(2021)Yao, Wang, Li, Sheng, Huo, Xi, Yang, and Zhang]{yao}
Mingjia Yao, Yuxiang Wang, Xin Li, Ye~Sheng, Haiyang Huo, Lili Xi, Jiong Yang,
  and Wenqing Zhang.
\newblock Materials informatics platform with three dimensional structures,
  workflow and thermoelectric applications.
\newblock \emph{Scientific Data}, 8\penalty0 (1):\penalty0 236, 2021.
\newblock \doi{10.1038/s41597-021-01022-6}.
\newblock URL \url{https://doi.org/10.1038/s41597-021-01022-6}.

\bibitem[Miyazaki et~al.(2021)Miyazaki, Tamura, Mikami, Watanabe, Ide,
  Ozkendir, and Nishino]{miyazaki}
Hidetoshi Miyazaki, Tomoyuki Tamura, Masashi Mikami, Kosuke Watanabe, Naoki
  Ide, Osman~Murat Ozkendir, and Yoichi Nishino.
\newblock Machine learning based prediction of lattice thermal conductivity for
  half-heusler compounds using atomic information.
\newblock \emph{Scientific Reports}, 11\penalty0 (1):\penalty0 13410, 2021.
\newblock \doi{10.1038/s41598-021-92030-4}.
\newblock URL \url{https://doi.org/10.1038/s41598-021-92030-4}.

\bibitem[Yukari~Katsura and Tsuda(2019)]{katsura}
Takushi Kodani Mitsunori Kaneshige Yuki Ando Sakiko Gunji Yoji Imai Hideyasu
  Ouchi Kazuki Tobita Kaoru~Kimura Yukari~Katsura, Masaya~Kumagai and Koji
  Tsuda.
\newblock Data-driven analysis of electron relaxation times in pbte-type
  thermoelectric materials.
\newblock \emph{Science and Technology of Advanced Materials}, 20\penalty0
  (1):\penalty0 511--520, 2019.
\newblock \doi{10.1080/14686996.2019.1603885}.

\bibitem[Priya and Aluru(2021)]{priya}
Pikee Priya and N.~R. Aluru.
\newblock Accelerated design and discovery of perovskites with high
  conductivity for energy applications through machine learning.
\newblock \emph{npj Computational Materials}, 7\penalty0 (1):\penalty0 90,
  2021.
\newblock \doi{10.1038/s41524-021-00551-3}.
\newblock URL \url{https://doi.org/10.1038/s41524-021-00551-3}.

\bibitem[Na and Chang(2022)]{public_thermo}
Gyoung~S. Na and Hyunju Chang.
\newblock A public database of thermoelectric materials and system-identified
  material representation for data-driven discovery.
\newblock \emph{npj Computational Materials}, 8\penalty0 (1):\penalty0 214,
  2022.
\newblock \doi{10.1038/s41524-022-00897-2}.
\newblock URL \url{https://doi.org/10.1038/s41524-022-00897-2}.

\bibitem[Curtarolo et~al.(2012)Curtarolo, Setyawan, Hart, Jahnatek, Chepulskii,
  Taylor, Wang, Xue, Yang, Levy, Mehl, Stokes, Demchenko, and Morgan]{aflow}
Stefano Curtarolo, Wahyu Setyawan, Gus~L.W. Hart, Michal Jahnatek, Roman~V.
  Chepulskii, Richard~H. Taylor, Shidong Wang, Junkai Xue, Kesong Yang, Ohad
  Levy, Michael~J. Mehl, Harold~T. Stokes, Denis~O. Demchenko, and Dane Morgan.
\newblock Aflow: An automatic framework for high-throughput materials
  discovery.
\newblock \emph{Computational Materials Science}, 58:\penalty0 218--226, 2012.
\newblock ISSN 0927-0256.
\newblock \doi{https://doi.org/10.1016/j.commatsci.2012.02.005}.
\newblock URL
  \url{https://www.sciencedirect.com/science/article/pii/S0927025612000717}.

\bibitem[Cohen et~al.(2008)Cohen, Mori-Sánchez, and Yang]{cohen2008dft}
Aron~J. Cohen, Paula Mori-Sánchez, and Weitao Yang.
\newblock Insights into current limitations of density functional theory.
\newblock \emph{Science}, 321\penalty0 (5890):\penalty0 792--794, 2008.
\newblock \doi{10.1126/science.1158722}.
\newblock URL \url{https://www.science.org/doi/abs/10.1126/science.1158722}.

\bibitem[Gordon(2000)]{gordon2000}
Roy~G. Gordon.
\newblock Criteria for choosing transparent conductors.
\newblock \emph{MRS Bulletin}, 25\penalty0 (8):\penalty0 52–57, 2000.
\newblock \doi{10.1557/mrs2000.151}.

\bibitem[Clymo et~al.(2024)Clymo, Collins, Atkinson, Dyer, Gaultois, Gusev, and
  et~al.]{clymo2024comgen}
J.~Clymo, C.~M. Collins, K.~Atkinson, M.~S. Dyer, M.~W. Gaultois, V.~V. Gusev,
  and et~al.
\newblock Exploration of chemical space through automated reasoning.
\newblock \emph{ChemRxiv}, 2024.
\newblock \doi{10.26434/chemrxiv-2024-2bhfw}.
\newblock This content is a preprint and has not been peer-reviewed.

\bibitem[Mott(1985)]{mott1985metal}
N.F. Mott.
\newblock Is there ever a minimum metallic conductivity?
\newblock \emph{Solid-State Electronics}, 28\penalty0 (1):\penalty0 57--59,
  1985.
\newblock ISSN 0038-1101.
\newblock \doi{https://doi.org/10.1016/0038-1101(85)90210-2}.
\newblock URL
  \url{https://www.sciencedirect.com/science/article/pii/0038110185902102}.

\bibitem[Chudnovskii(1978)]{chudnovskii1978metal}
F~A Chudnovskii.
\newblock The minimum conductivity and electron localisation in the metallic
  phase of transition metal compounds in the vicinity of a metal-insulator
  transition.
\newblock \emph{Journal of Physics C: Solid State Physics}, 11\penalty0
  (3):\penalty0 L99, feb 1978.
\newblock \doi{10.1088/0022-3719/11/3/003}.
\newblock URL \url{https://dx.doi.org/10.1088/0022-3719/11/3/003}.

\bibitem[Dunn et~al.(2020)Dunn, Wang, Ganose, Dopp, and Jain]{matbench}
Alexander Dunn, Qi~Wang, Alex Ganose, Daniel Dopp, and Anubhav Jain.
\newblock Benchmarking materials property prediction methods: the matbench test
  set and automatminer reference algorithm.
\newblock \emph{npj Computational Materials}, 6\penalty0 (1):\penalty0 138,
  2020.
\newblock \doi{10.1038/s41524-020-00406-3}.
\newblock URL \url{https://doi.org/10.1038/s41524-020-00406-3}.

\bibitem[Sivakumar et~al.(2021{\natexlab{a}})Sivakumar, Akkera, {Ranjeth Kumar
  Reddy}, {Srinivas Reddy}, Kambhala, and {Nanda Kumar Reddy}]{sno2:ga}
Peddavarapu Sivakumar, Harish~Sharma Akkera, T.~{Ranjeth Kumar Reddy},
  G.~{Srinivas Reddy}, Nagaiah Kambhala, and N.~{Nanda Kumar Reddy}.
\newblock Influence of $\text{Ga}$ doping on structural, optical and electrical
  properties of transparent conducting $\text{SnO}_2$ thin films.
\newblock \emph{Optik}, 226:\penalty0 165859, 2021{\natexlab{a}}.
\newblock ISSN 0030-4026.
\newblock \doi{https://doi.org/10.1016/j.ijleo.2020.165859}.
\newblock URL
  \url{https://www.sciencedirect.com/science/article/pii/S0030402620316764}.

\bibitem[Teldja et~al.(2020)Teldja, Noureddine, Azzeddine, and Meriem]{sno2:in}
Boucherka Teldja, Brihi Noureddine, Berbadj Azzeddine, and Touati Meriem.
\newblock Effect of indium doping on the uv photoluminescence emission,
  structural, electrical and optical properties of spin-coating deposited
  $\text{SnO}_2$ thin films.
\newblock \emph{Optik}, 209:\penalty0 164586, 2020.
\newblock ISSN 0030-4026.
\newblock \doi{https://doi.org/10.1016/j.ijleo.2020.164586}.
\newblock URL
  \url{https://www.sciencedirect.com/science/article/pii/S0030402620304204}.

\bibitem[Arora et~al.(2021)Arora, Malhotra, Mahajan, and Kumar]{sno2:mn}
Isha Arora, Kamini Malhotra, Alish Mahajan, and Praveen Kumar.
\newblock Structural, optical and electrical characterization of spin coated
  $\text{SnO}_{2}$:$\text{Mn}$ thin films.
\newblock \emph{Materials Today: Proceedings}, 36:\penalty0 697--700, 2021.
\newblock ISSN 2214-7853.
\newblock \doi{https://doi.org/10.1016/j.matpr.2020.04.750}.
\newblock URL
  \url{https://www.sciencedirect.com/science/article/pii/S2214785320333988}.
\newblock National Conference on Advanced Functional Materials 2019.

\bibitem[Uwihoreye et~al.(2023)Uwihoreye, Yang, Zhang, Lin, Liang, Yang, and
  Zhang]{sno2:ta}
Vedaste Uwihoreye, Zhenni Yang, Jia-Ye Zhang, Yu-Mei Lin, Xuan Liang, Lu~Yang,
  and Kelvin H.~L. Zhang.
\newblock Transparent conductive $\text{SnO}_{2}$ thin films via resonant
  $\text{Ta}$ doping.
\newblock \emph{Science China Materials}, 66\penalty0 (1):\penalty0 264--271,
  2023.
\newblock \doi{10.1007/s40843-022-2122-9}.
\newblock URL \url{https://doi.org/10.1007/s40843-022-2122-9}.

\bibitem[Sivakumar et~al.(2021{\natexlab{b}})Sivakumar, Akkera, {Kumar Reddy},
  Bitla, Ganesh, Kumar, Reddy, and Poloju]{sno2:ti}
Peddavarapu Sivakumar, Harish~Sharma Akkera, T.~Ranjeth {Kumar Reddy},
  Yugandhar Bitla, V.~Ganesh, P.~Mohan Kumar, G.~Srinivas Reddy, and Madhukar
  Poloju.
\newblock Effect of $\text{Ti}$ doping on structural, optical and electrical
  properties of $\text{SnO}_{2}$ transparent conducting thin films deposited by
  sol-gel spin coating.
\newblock \emph{Optical Materials}, 113:\penalty0 110845, 2021{\natexlab{b}}.
\newblock ISSN 0925-3467.
\newblock \doi{https://doi.org/10.1016/j.optmat.2021.110845}.
\newblock URL
  \url{https://www.sciencedirect.com/science/article/pii/S092534672100046X}.

\bibitem[Wang et~al.(2013)Wang, Gao, Chen, Cao, Zhou, Dai, and Guo]{sno2:w}
Mi~Wang, Yanfeng Gao, Zhang Chen, Chuanxiang Cao, Jiadong Zhou, Lei Dai, and
  Xuhong Guo.
\newblock Transparent and conductive $\text{W}$-doped $\text{SnO}_{\text{2}}$
  thin films fabricated by an aqueous solution process.
\newblock \emph{Thin Solid Films}, 544:\penalty0 419--426, 2013.
\newblock ISSN 0040-6090.
\newblock \doi{https://doi.org/10.1016/j.tsf.2013.02.088}.
\newblock URL
  \url{https://www.sciencedirect.com/science/article/pii/S0040609013003581}.
\newblock The 6th International Conference on Technological Advances of Thin
  Films and Surface Coatings.

\bibitem[Shen et~al.(2017)Shen, An, Zhang, Zhang, Wu, Yan, and Liu]{in2o3_sn_1}
Luhang Shen, Yukai An, Rukang Zhang, Pan Zhang, Zhonghua Wu, Hui Yan, and Jiwen
  Liu.
\newblock Enhanced room-temperature ferromagnetism on $(\text{In}_{0.98-x}
  \text{Co}_{x} \text{Sn}_{0.02})_{2}\text{O}_{3}$ films: magnetic mechanism,
  optical and transport properties.
\newblock \emph{Phys. Chem. Chem. Phys.}, 19:\penalty0 29472--29482, 2017.
\newblock \doi{10.1039/C7CP05764D}.
\newblock URL \url{http://dx.doi.org/10.1039/C7CP05764D}.

\bibitem[Fallah et~al.(2007)Fallah, Ghasemi, Hassanzadeh, and
  Steki]{in2o3_sn_2}
Hamid~Reza Fallah, Mohsen Ghasemi, Ali Hassanzadeh, and Hadi Steki.
\newblock The effect of annealing on structural, electrical and optical
  properties of nanostructured $\text{ITO}$ films prepared by e-beam
  evaporation.
\newblock \emph{Materials Research Bulletin}, 42\penalty0 (3):\penalty0
  487--496, 2007.
\newblock ISSN 0025-5408.
\newblock \doi{https://doi.org/10.1016/j.materresbull.2006.06.024}.
\newblock URL
  \url{https://www.sciencedirect.com/science/article/pii/S002554080600273X}.

\bibitem[Ambrosini et~al.(2000)Ambrosini, Duarte, Poeppelmeier, Lane,
  Kannewurf, and Mason]{in2o3_sn_3}
Andrea Ambrosini, Angel Duarte, Kenneth~R. Poeppelmeier, Melissa Lane, Carl~R.
  Kannewurf, and Thomas~O. Mason.
\newblock Electrical, optical, and structural properties of tin-doped
  $\text{In}_{2}\text{O}_{3}$–$\text{M}_{2}\text{O}_{3}$ solid solutions
  ($\text{M=Y}$, $\text{Sc}$).
\newblock \emph{Journal of Solid State Chemistry}, 153\penalty0 (1):\penalty0
  41--47, 2000.
\newblock ISSN 0022-4596.
\newblock \doi{https://doi.org/10.1006/jssc.2000.8737}.
\newblock URL
  \url{https://www.sciencedirect.com/science/article/pii/S0022459600987371}.

\bibitem[Guendouz et~al.(2018)Guendouz, Bouaine, and Brihi]{zno:al-sn}
Hassan Guendouz, Abdelhamid Bouaine, and Noureddine Brihi.
\newblock Biphase effect on structural, optical, and electrical properties of
  $\text{Al-Sn}$ codoped $\text{ZnO}$ thin films deposited by sol-gel
  spin-coating technique.
\newblock \emph{Optik}, 158:\penalty0 1342--1348, 2018.
\newblock ISSN 0030-4026.
\newblock \doi{https://doi.org/10.1016/j.ijleo.2018.01.025}.
\newblock URL
  \url{https://www.sciencedirect.com/science/article/pii/S0030402618300287}.

\bibitem[Shukla et~al.(2006)Shukla, Srivastava, Srivastava, and Dubey]{zno:al}
R.K. Shukla, Anchal Srivastava, Atul Srivastava, and K.C. Dubey.
\newblock Growth of transparent conducting nanocrystalline $\text{Al}$ doped
  $\text{ZnO}$ thin films by pulsed laser deposition.
\newblock \emph{Journal of Crystal Growth}, 294\penalty0 (2):\penalty0
  427--431, 2006.
\newblock ISSN 0022-0248.
\newblock \doi{https://doi.org/10.1016/j.jcrysgro.2006.06.035}.
\newblock URL
  \url{https://www.sciencedirect.com/science/article/pii/S0022024806006178}.

\bibitem[Ajimsha et~al.(2015)Ajimsha, Das, Misra, Joshi, Kukreja, Kumar,
  Sharma, and Oak]{zno:ga}
R.S. Ajimsha, Amit~K. Das, P.~Misra, M.P. Joshi, L.M. Kukreja, R.~Kumar, T.K.
  Sharma, and S.M. Oak.
\newblock Observation of low resistivity and high mobility in $\text{Ga}$ doped
  $\text{ZnO}$ thin films grown by buffer assisted pulsed laser deposition.
\newblock \emph{Journal of Alloys and Compounds}, 638:\penalty0 55--58, 2015.
\newblock ISSN 0925-8388.
\newblock \doi{https://doi.org/10.1016/j.jallcom.2015.02.162}.
\newblock URL
  \url{https://www.sciencedirect.com/science/article/pii/S0925838815006027}.

\bibitem[Breiman(2001)]{random_forest}
Leo Breiman.
\newblock Random forests.
\newblock \emph{Machine Learning}, 45\penalty0 (1):\penalty0 5--32, 2001.
\newblock \doi{10.1023/A:1010933404324}.
\newblock URL \url{https://doi.org/10.1023/A:1010933404324}.

\bibitem[Venkatraman(2021)]{rf_gap_cbfv}
Vishwesh Venkatraman.
\newblock The utility of composition-based machine learning models for band gap
  prediction.
\newblock \emph{Computational Materials Science}, 197:\penalty0 110637, 2021.
\newblock ISSN 0927-0256.
\newblock \doi{https://doi.org/10.1016/j.commatsci.2021.110637}.
\newblock URL
  \url{https://www.sciencedirect.com/science/article/pii/S0927025621003645}.

\bibitem[Riebesell(2019)]{riebesell_thermoelectric}
Janosh Riebesell.
\newblock {Probabilistic Data-Driven Discovery of Thermoelectric Materials}.
\newblock \emph{MPhil thesis, University of Cambridge}, 2019.
\newblock URL \url{https://github.com/janosh/thermo}.

\bibitem[Chelladurai et~al.(2022)Chelladurai, Upreti, Verma, Agrawal, Garg,
  Kaushik, Agrawal, Singh, and Narayanasamy]{rf_mech_strength}
Samson Jerold~Samuel Chelladurai, Kamal Upreti, Manvendra Verma, Meena Agrawal,
  Jatinder Garg, Rekha Kaushik, Chinmay Agrawal, Divakar Singh, and Rajamani
  Narayanasamy.
\newblock Prediction of mechanical strength by using an artificial neural
  network and random forest algorithm.
\newblock \emph{Journal of Nanomaterials}, 2022:\penalty0 7791582, 2022.
\newblock \doi{10.1155/2022/7791582}.
\newblock URL \url{https://doi.org/10.1155/2022/7791582}.

\bibitem[Ward et~al.(2016)Ward, Agrawal, Choudhary, and Wolverton]{magpie}
Logan Ward, Ankit Agrawal, Alok Choudhary, and Christopher Wolverton.
\newblock A general-purpose machine learning framework for predicting
  properties of inorganic materials.
\newblock \emph{npj Computational Materials}, 2\penalty0 (1):\penalty0 16028,
  Aug 2016.
\newblock ISSN 2057-3960.
\newblock \doi{10.1038/npjcompumats.2016.28}.
\newblock URL \url{https://doi.org/10.1038/npjcompumats.2016.28}.

\bibitem[Sonpal et~al.(2022)Sonpal, Afzal, An, Chandrasekaran, and
  Halls]{cryst_descriptors}
Aditya Sonpal, Mohammad Atif~Faiz Afzal, Yuling An, Anand Chandrasekaran, and
  Mathew~D. Halls.
\newblock \emph{Benchmarking Machine Learning Descriptors for Crystals}, volume
  1416 of \emph{ACS Symposium Series}, doi:10.1021/bk-2022-1416.ch006~6, pages
  111--126.
\newblock American Chemical Society, 2023/02/14 2022.
\newblock ISBN 9780841297630.
\newblock \doi{doi:10.1021/bk-2022-1416.ch006}.
\newblock URL \url{https://doi.org/10.1021/bk-2022-1416.ch006}.

\bibitem[Wang et~al.(2021)Wang, Kauwe, Murdock, and Sparks]{crabnet}
Anthony Yu-Tung Wang, Steven~K. Kauwe, Ryan~J. Murdock, and Taylor~D. Sparks.
\newblock Compositionally restricted attention-based network for materials
  property predictions.
\newblock \emph{npj Computational Materials}, 7\penalty0 (1):\penalty0 77, May
  2021.
\newblock ISSN 2057-3960.
\newblock \doi{10.1038/s41524-021-00545-1}.
\newblock URL \url{https://doi.org/10.1038/s41524-021-00545-1}.

\bibitem[Vaswani et~al.(2017{\natexlab{a}})Vaswani, Shazeer, Parmar, Uszkoreit,
  Jones, Gomez, Kaiser, and Polosukhin]{transformers}
Ashish Vaswani, Noam Shazeer, Niki Parmar, Jakob Uszkoreit, Llion Jones,
  Aidan~N. Gomez, Lukasz Kaiser, and Illia Polosukhin.
\newblock Attention is all you need, 2017{\natexlab{a}}.
\newblock URL \url{https://arxiv.org/abs/1706.03762}.

\bibitem[Baird et~al.(2022)Baird, Diep, and Sparks]{discover}
Sterling~G. Baird, Tran~Q. Diep, and Taylor~D. Sparks.
\newblock Discover: a materials discovery screening tool for high
  performance{,} unique chemical compositions.
\newblock \emph{Digital Discovery}, 1:\penalty0 226--240, 2022.
\newblock \doi{10.1039/D1DD00028D}.

\bibitem[Hargreaves et~al.(2023)Hargreaves, Gaultois, Daniels, Watts, Kurlin,
  Moran, Dang, Morris, Morscher, Thompson, Wright, Prasad, Blanc, Collins,
  Crawford, Duff, Evans, Gamon, Han, Leube, Niu, Perez, Robinson, Rogan, Sharp,
  Shoko, Sonni, Thomas, Vasylenko, Wang, Rosseinsky, and Dyer]{hargreaves_2023}
Cameron~J. Hargreaves, Michael~W. Gaultois, Luke~M. Daniels, Emma~J. Watts,
  Vitaliy~A. Kurlin, Michael Moran, Yun Dang, Rhun Morris, Alexandra Morscher,
  Kate Thompson, Matthew~A. Wright, Beluvalli-Eshwarappa Prasad,
  Fr{\'e}d{\'e}ric Blanc, Chris~M. Collins, Catriona~A. Crawford, Benjamin~B.
  Duff, Jae Evans, Jacinthe Gamon, Guopeng Han, Bernhard~T. Leube, Hongjun Niu,
  Arnaud~J. Perez, Aris Robinson, Oliver Rogan, Paul~M. Sharp, Elvis Shoko,
  Manel Sonni, William~J. Thomas, Andrij Vasylenko, Lu~Wang, Matthew~J.
  Rosseinsky, and Matthew~S. Dyer.
\newblock A database of experimentally measured lithium solid electrolyte
  conductivities evaluated with machine learning.
\newblock \emph{npj Computational Materials}, 9\penalty0 (1):\penalty0 9, 2023.
\newblock \doi{10.1038/s41524-022-00951-z}.
\newblock URL \url{https://doi.org/10.1038/s41524-022-00951-z}.

\bibitem[Antunes et~al.(2023)Antunes, Butler, and Grau-Crespo]{cratenet}
Luis~M Antunes, Keith~T Butler, and Ricardo Grau-Crespo.
\newblock Predicting thermoelectric transport properties from composition with
  attention-based deep learning.
\newblock \emph{Machine Learning: Science and Technology}, 4\penalty0
  (1):\penalty0 015037, apr 2023.
\newblock \doi{10.1088/2632-2153/acc4a9}.
\newblock URL \url{https://dx.doi.org/10.1088/2632-2153/acc4a9}.

\bibitem[Lee et~al.(2023{\natexlab{a}})Lee, Park, Yang, Han, and
  Lim]{lee2023clcs}
Jaewan Lee, Changyoung Park, Hongjun Yang, Sehui Han, and Woohyung Lim.
\newblock {CLCS} : Contrastive learning between compositions and structures for
  practical $\text{Li}$-ion battery electrodes design.
\newblock In \emph{AI for Accelerated Materials Design - NeurIPS 2023
  Workshop}, 2023{\natexlab{a}}.
\newblock URL \url{https://openreview.net/forum?id=FfvByyoVAO}.

\bibitem[Mukherjee et~al.(2020{\natexlab{b}})Mukherjee, Satsangi, and
  Singh]{mukherjee_sigma}
Madhubanti Mukherjee, Swanti Satsangi, and Abhishek~K. Singh.
\newblock A statistical approach for the rapid prediction of electron
  relaxation time using elemental representatives, Jul 2020{\natexlab{b}}.
\newblock URL
  \url{https://acs.figshare.com/collections/A_Statistical_Approach_for_the_Rapid_Prediction_of_Electron_Relaxation_Time_Using_Elemental_Representatives/5071658/1}.

\bibitem[Nguyen-Sy et~al.(2021)Nguyen-Sy, To, Vu, Nguyen, and
  Nguyen]{nguyen_sigma}
Tuan Nguyen-Sy, Quy-Dong To, Minh-Ngoc Vu, The-Duong Nguyen, and Thoi-Trung
  Nguyen.
\newblock Predicting the electrical conductivity of brine-saturated rocks using
  machine learning methods.
\newblock \emph{Journal of Applied Geophysics}, 184:\penalty0 104238, 2021.
\newblock ISSN 0926-9851.
\newblock \doi{https://doi.org/10.1016/j.jappgeo.2020.104238}.
\newblock URL
  \url{https://www.sciencedirect.com/science/article/pii/S0926985120306212}.

\bibitem[Hastie et~al.(2001)Hastie, Tibshirani, and Friedman]{tibshirani_2001}
Trevor Hastie, Robert Tibshirani, and Jerome Friedman.
\newblock \emph{The Elements of Statistical Learning}.
\newblock Springer Series in Statistics. Springer New York Inc., New York, NY,
  USA, 2001.

\bibitem[Meredig et~al.(2018)Meredig, Antono, Church, Hutchinson, Ling,
  Paradiso, Blaiszik, Foster, Gibbons, Hattrick-Simpers, Mehta, and
  Ward]{meredig_2018}
Bryce Meredig, Erin Antono, Carena Church, Maxwell Hutchinson, Julia Ling, Sean
  Paradiso, Ben Blaiszik, Ian Foster, Brenna Gibbons, Jason Hattrick-Simpers,
  Apurva Mehta, and Logan Ward.
\newblock Can machine learning identify the next high-temperature
  superconductor? examining extrapolation performance for materials discovery.
\newblock \emph{Mol. Syst. Des. Eng.}, 3:\penalty0 819--825, 2018.
\newblock \doi{10.1039/C8ME00012C}.
\newblock URL \url{http://dx.doi.org/10.1039/C8ME00012C}.

\bibitem[Ottomano et~al.(2024)Ottomano, De~Felice, Gusev, and Sparks]{ottomano}
Federico Ottomano, Giovanni De~Felice, Vladimir~V. Gusev, and Taylor~D. Sparks.
\newblock Not as simple as we thought: a rigorous examination of data
  aggregation in materials informatics.
\newblock \emph{Digital Discovery}, pages~--, 2024.
\newblock \doi{10.1039/D3DD00207A}.
\newblock URL \url{http://dx.doi.org/10.1039/D3DD00207A}.

\bibitem[Li et~al.(2023)Li, DeCost, Choudhary, Greenwood, and
  Hattrick-Simpers]{critical_ex_gen}
Kangming Li, Brian DeCost, Kamal Choudhary, Michael Greenwood, and Jason
  Hattrick-Simpers.
\newblock A critical examination of robustness and generalizability of machine
  learning prediction of materials properties.
\newblock \emph{npj Computational Materials}, 9\penalty0 (1):\penalty0 55,
  2023.
\newblock \doi{10.1038/s41524-023-01012-9}.
\newblock URL \url{https://doi.org/10.1038/s41524-023-01012-9}.

\bibitem[Omee et~al.(2024)Omee, Fu, Dong, Hu, and Hu]{omee2024structurebased}
Sadman~Sadeed Omee, Nihang Fu, Rongzhi Dong, Ming Hu, and Jianjun Hu.
\newblock Structure-based out-of-distribution (ood) materials property
  prediction: a benchmark study, 2024.

\bibitem[Lloyd(1982)]{K-means}
Stuart~P. Lloyd.
\newblock Least squares quantization in pcm.
\newblock \emph{IEEE Trans. Inf. Theory}, 28:\penalty0 129--136, 1982.

\bibitem[Durdy et~al.(2022)Durdy, Gaultois, Gusev, Bollegala, and
  Rosseinsky]{random_proj_sam}
Samantha Durdy, Michael~W. Gaultois, Vladimir~V. Gusev, Danushka Bollegala, and
  Matthew~J. Rosseinsky.
\newblock Random projections and kernelised leave one cluster out cross
  validation: universal baselines and evaluation tools for supervised machine
  learning of material properties.
\newblock \emph{Digital Discovery}, 1:\penalty0 763--778, 2022.
\newblock \doi{10.1039/D2DD00039C}.
\newblock URL \url{http://dx.doi.org/10.1039/D2DD00039C}.

\bibitem[Hofmann et~al.(2008)Hofmann, Sch{\"o}lkopf, and Smola]{kernels}
Thomas Hofmann, Bernhard Sch{\"o}lkopf, and Alexander~J. Smola.
\newblock {Kernel methods in machine learning}.
\newblock \emph{The Annals of Statistics}, 36\penalty0 (3):\penalty0 1171 --
  1220, 2008.
\newblock \doi{10.1214/009053607000000677}.
\newblock URL \url{https://doi.org/10.1214/009053607000000677}.

\bibitem[Wang et~al.(2004)Wang, Chen, and Chen]{rbf}
Junping Wang, Quanshi Chen, and Yong Chen.
\newblock $\text{RBF}$ kernel based support vector machine with universal
  approximation and its application.
\newblock In Fu-Liang Yin, Jun Wang, and Chengan Guo, editors, \emph{Advances
  in Neural Networks -- ISNN 2004}, pages 512--517, Berlin, Heidelberg, 2004.
  Springer Berlin Heidelberg.
\newblock ISBN 978-3-540-28647-9.

\bibitem[Shoghi et~al.(2023)Shoghi, Kolluru, Kitchin, Ulissi, Zitnick, and
  Wood]{shoghi2023molecules}
Nima Shoghi, Adeesh Kolluru, John~R. Kitchin, Zachary~W. Ulissi, C.~Lawrence
  Zitnick, and Brandon~M. Wood.
\newblock From molecules to materials: Pre-training large generalizable models
  for atomic property prediction, 2023.

\bibitem[Jablonka et~al.(2024)Jablonka, Schwaller, Ortega-Guerrero, and
  Smit]{jablonka2024}
Kevin Jablonka, Philippe Schwaller, Andrés Ortega-Guerrero, and Berend Smit.
\newblock Leveraging large language models for predictive chemistry.
\newblock \emph{Nature Machine Intelligence}, 6:\penalty0 1--9, 02 2024.
\newblock \doi{10.1038/s42256-023-00788-1}.

\bibitem[Chicco and Jurman(2020)]{chicco2020mcc}
Davide Chicco and Giuseppe Jurman.
\newblock The advantages of the matthews correlation coefficient (mcc) over f1
  score and accuracy in binary classification evaluation.
\newblock \emph{BMC Genomics}, 21\penalty0 (1):\penalty0 6, 2020.
\newblock \doi{10.1186/s12864-019-6413-7}.
\newblock URL \url{https://doi.org/10.1186/s12864-019-6413-7}.

\bibitem[Linardatos et~al.(2021)Linardatos, Papastefanopoulos, and
  Kotsiantis]{xai2021linardatos}
Pantelis Linardatos, Vasilis Papastefanopoulos, and Sotiris Kotsiantis.
\newblock Explainable $\text{AI}$: A review of machine learning
  interpretability methods.
\newblock \emph{Entropy}, 23\penalty0 (1), 2021.
\newblock ISSN 1099-4300.
\newblock \doi{10.3390/e23010018}.
\newblock URL \url{https://www.mdpi.com/1099-4300/23/1/18}.

\bibitem[Vaswani et~al.(2017{\natexlab{b}})Vaswani, Shazeer, Parmar, Uszkoreit,
  Jones, Gomez, Kaiser, and Polosukhin]{vaswani}
Ashish Vaswani, Noam Shazeer, Niki Parmar, Jakob Uszkoreit, Llion Jones,
  Aidan~N Gomez, \L~ukasz Kaiser, and Illia Polosukhin.
\newblock Attention is all you need.
\newblock In I.~Guyon, U.~Von Luxburg, S.~Bengio, H.~Wallach, R.~Fergus,
  S.~Vishwanathan, and R.~Garnett, editors, \emph{Advances in Neural
  Information Processing Systems}, volume~30. Curran Associates, Inc.,
  2017{\natexlab{b}}.
\newblock URL
  \url{https://proceedings.neurips.cc/paper/2017/file/3f5ee243547dee91fbd053c1c4a845aa-Paper.pdf}.

\bibitem[Erdogan et~al.(2021)Erdogan, Kutlu, Sedefoglu, and
  Kavak]{erdogan2021effect}
NH~Erdogan, T~Kutlu, N~Sedefoglu, and HAM{\.I}DE Kavak.
\newblock Effect of $\text{Na}$ doping on microstructures, optical and
  electrical properties of $\text{ZnO}$ thin films grown by sol-gel method.
\newblock \emph{Journal of Alloys and Compounds}, 881:\penalty0 160554, 2021.

\bibitem[Basyooni et~al.(2017)Basyooni, Shaban, and
  El~Sayed]{basyooni2017enhanced}
Mohamed~A Basyooni, Mohamed Shaban, and Adel~M El~Sayed.
\newblock Enhanced gas sensing properties of spin-coated $\text{Na}$-doped
  $\text{ZnO}$ nanostructured films.
\newblock \emph{Scientific reports}, 7\penalty0 (1):\penalty0 41716, 2017.

\bibitem[Mahdhi et~al.(2018)Mahdhi, Djessas, and Ayadi]{mahdhi2018synthesis}
H~Mahdhi, K~Djessas, and Z~Ben Ayadi.
\newblock Synthesis and characteristics of $\text{Ca}$-doped $\text{ZnO}$ thin
  films by rf magnetron sputtering at low temperature.
\newblock \emph{Materials Letters}, 214:\penalty0 10--14, 2018.

\bibitem[Edwards and Mason(1998)]{edwards1998subsolidus}
Doreen~D Edwards and Thomas~O Mason.
\newblock Subsolidus phase relations in the
  $\text{Ga}_2\text{O}_3$-$\text{In}_2\text{O}_3$-$\text{SnO}_2$ system.
\newblock \emph{Journal of the American Ceramic Society}, 81\penalty0
  (12):\penalty0 3285--3292, 1998.

\bibitem[Edwards et~al.(1997)Edwards, Mason, Goutenoire, and
  Poeppelmeier]{edwards1997new}
a~DD Edwards, TO~Mason, F~Goutenoire, and KR~Poeppelmeier.
\newblock A new transparent conducting oxide in the
  $\text{Ga}_2\text{O}_3$-$\text{In}_2\text{O}_3$-$\text{SnO}_2$ system.
\newblock \emph{Applied physics letters}, 70\penalty0 (13):\penalty0
  1706--1708, 1997.

\bibitem[Dolgonos et~al.(2015)Dolgonos, Wells, Poeppelmeier, and
  Mason]{dolgonos2015phase}
Alex Dolgonos, Spencer~A Wells, Kenneth~R Poeppelmeier, and Thomas~O Mason.
\newblock Phase stability and optoelectronic properties of the bixbyite phase
  in the gallium--indium--tin--oxide system.
\newblock \emph{Journal of the American Ceramic Society}, 98\penalty0
  (2):\penalty0 669--674, 2015.

\bibitem[Michiue et~al.(2020)Michiue, Son, and Mori]{michiue2020utilizing}
Yuichi Michiue, H-W Son, and Takao Mori.
\newblock Utilizing a unified structure model in (3+1)-dimensional superspace
  to identify a homologous phase
  \((\text{Ga}_{1-\alpha}\text{Al}_{\alpha})_{2}\text{O}_3(\text{ZnO})_{\text{m}}\)
  in $\text{ZnO}$-based thermoelectric composites.
\newblock \emph{Journal of Applied Crystallography}, 53\penalty0 (6):\penalty0
  1542--1549, 2020.

\bibitem[Jung et~al.(2020)Jung, Park, Gedi, Reddy, Ferblantier, and
  Kim]{jung2020doped}
Hyunmin Jung, Youngsang Park, Sreedevi Gedi, Vasudeva Reddy~Minnam Reddy,
  G{\'e}rald Ferblantier, and Woo~Kyoung Kim.
\newblock $\text{Al}$-doped zinc stannate films for photovoltaic applications.
\newblock \emph{Korean Journal of Chemical Engineering}, 37:\penalty0 730--735,
  2020.

\bibitem[Orita et~al.(2000)Orita, Tanji, Mizuno, Adachi, and
  Tanaka]{orita2000mechanism}
Masahiro Orita, Hiroaki Tanji, Masataka Mizuno, Hirohiko Adachi, and Isao
  Tanaka.
\newblock Mechanism of electrical conductivity of transparent
  $\text{InGaZnO}_4$.
\newblock \emph{Physical Review B}, 61\penalty0 (3):\penalty0 1811, 2000.

\bibitem[Altynbek et~al.(2013)]{altynbek2013carrier}
Murat Altynbek et~al.
\newblock Carrier generation in multicomponent wide-bandgap oxides:
  $\text{InGaZnO}_4$.
\newblock 2013.

\bibitem[Kauwe et~al.(2020)Kauwe, Graser, Murdock, and Sparks]{ml_extrap_mats}
Steven~K. Kauwe, Jake Graser, Ryan Murdock, and Taylor~D. Sparks.
\newblock Can machine learning find extraordinary materials?
\newblock \emph{Computational Materials Science}, 174:\penalty0 109498, 2020.
\newblock ISSN 0927-0256.
\newblock \doi{https://doi.org/10.1016/j.commatsci.2019.109498}.
\newblock URL
  \url{https://www.sciencedirect.com/science/article/pii/S0927025619307979}.

\bibitem[Xie et~al.(2023)Xie, Wan, Zhou, Huang, Liu, Linghu, Wang, Kit,
  Grazian, Zhang, et~al.]{xie2023large}
Tong Xie, Yuwei Wan, Yufei Zhou, Wei Huang, Yixuan Liu, Qingyuan Linghu,
  Shaozhou Wang, Chunyu Kit, Clara Grazian, Wenjie Zhang, et~al.
\newblock Large language models as master key: unlocking the secrets of
  materials science.
\newblock \emph{Available at SSRN 4534137}, 2023.

\bibitem[Lee et~al.(2023{\natexlab{b}})Lee, Noh, Na, Fu, Sun, and
  Park]{lee2023stoichiometry}
Namkyeong Lee, Heewoong Noh, Gyoung~S Na, Tianfan Fu, Jimeng Sun, and Chanyoung
  Park.
\newblock Stoichiometry representation learning with polymorphic crystal
  structures.
\newblock \emph{arXiv preprint arXiv:2312.13289}, 2023{\natexlab{b}}.

\bibitem[Goodall and Lee(2020)]{Roost}
Rhys E.~A. Goodall and Alpha~A. Lee.
\newblock Predicting materials properties without crystal structure: deep
  representation learning from stoichiometry.
\newblock \emph{Nature Communications}, 11\penalty0 (1):\penalty0 6280, Dec
  2020.
\newblock ISSN 2041-1723.
\newblock \doi{10.1038/s41467-020-19964-7}.
\newblock URL \url{https://doi.org/10.1038/s41467-020-19964-7}.

\bibitem[Lakshminarayanan et~al.(2016)Lakshminarayanan, Pritzel, and
  Blundell]{deep_ensembles}
Balaji Lakshminarayanan, Alexander Pritzel, and Charles Blundell.
\newblock Simple and scalable predictive uncertainty estimation using deep
  ensembles, 2016.
\newblock URL \url{https://arxiv.org/abs/1612.01474}.

\bibitem[Arbel et~al.(2023)Arbel, Pitas, Vladimirova, and
  Fortuin]{arbel2023primerbnns}
Julyan Arbel, Konstantinos Pitas, Mariia Vladimirova, and Vincent Fortuin.
\newblock A primer on bayesian neural networks: Review and debates, 2023.
\newblock URL \url{https://arxiv.org/abs/2309.16314}.

\end{thebibliography}





\newpage
\section*{Appendix}
\appendix
\section{Uncertainty quantification}\label{app:unc_est}
Uncertainty quantification is critical for ML models in materials discovery, as experimental validation is resource-intensive. It is essential to model uncertainties in predictions to improve reliability and guide experimental efforts effectively.
Uncertainty is typically categorized into two types:
\begin{itemize}
    \item \textbf{Aleatoric uncertainty:} intrinsic noise in the observations, reducible only by improving data quality. It can be homoscedastic (constant variance) or heteroscedastic (variance dependent on specific inputs), with heteroscedastic uncertainty being common in materials science due to varying measurement conditions and sample qualities.

    \item \textbf{Epistemic uncertainty:} Model uncertainty due to insufficient data. It is reducible by incorporating more data in the training process.
\end{itemize}
Deep learning models do not naturally capture uncertainties, often yielding overconfident predictions. Aleatoric uncertainty in neural-network models can be captured by predicting the parameters of a heteroscedastic Gaussian distribution from the last layer, modeling both the predictive mean $f_{\theta}(\mathbf{x}_i)$ and variance $\hat{\sigma}_{a, \theta}^2(\mathbf{x}_i)$ using a \textit{Robust} loss function \citep{riebesell_thermoelectric, Roost}:
\begin{equation}
	\mathcal{L}(\theta) = \frac{1}{N}\sum_{i=1}^N \frac{|y_i - f_\theta(\bm{x}_i)|}{2\hat{\sigma}_{a,\theta}^2(\textbf{x}_i)} + \frac{1}{2}\text{log}(\hat{\sigma}_{a,\theta}^2(\textbf{x}_i)).
	\label{robust_loss}
\end{equation}
For epistemic contribution to the uncertainty, deep ensembles~\citep{deep_ensembles} are used, where the variance across predictions from multiple neural networks approximates the bayesian predictive distribution. The final uncertainty combines aleatoric and epistemic components:
\begin{equation}
	\hat{\sigma}^2(\bm{x}_i) = \hat{\sigma}_e^2(\bm{x}_i) + \frac{1}{M}\sum_{m=1}^M \hat{\sigma}_{a,\theta_m}^2(\bm{x}_i) \, ,
	\label{eq:total_unc}
\end{equation}
\begin{figure*}[!t]
\centering
\includegraphics[width=0.91\textwidth]{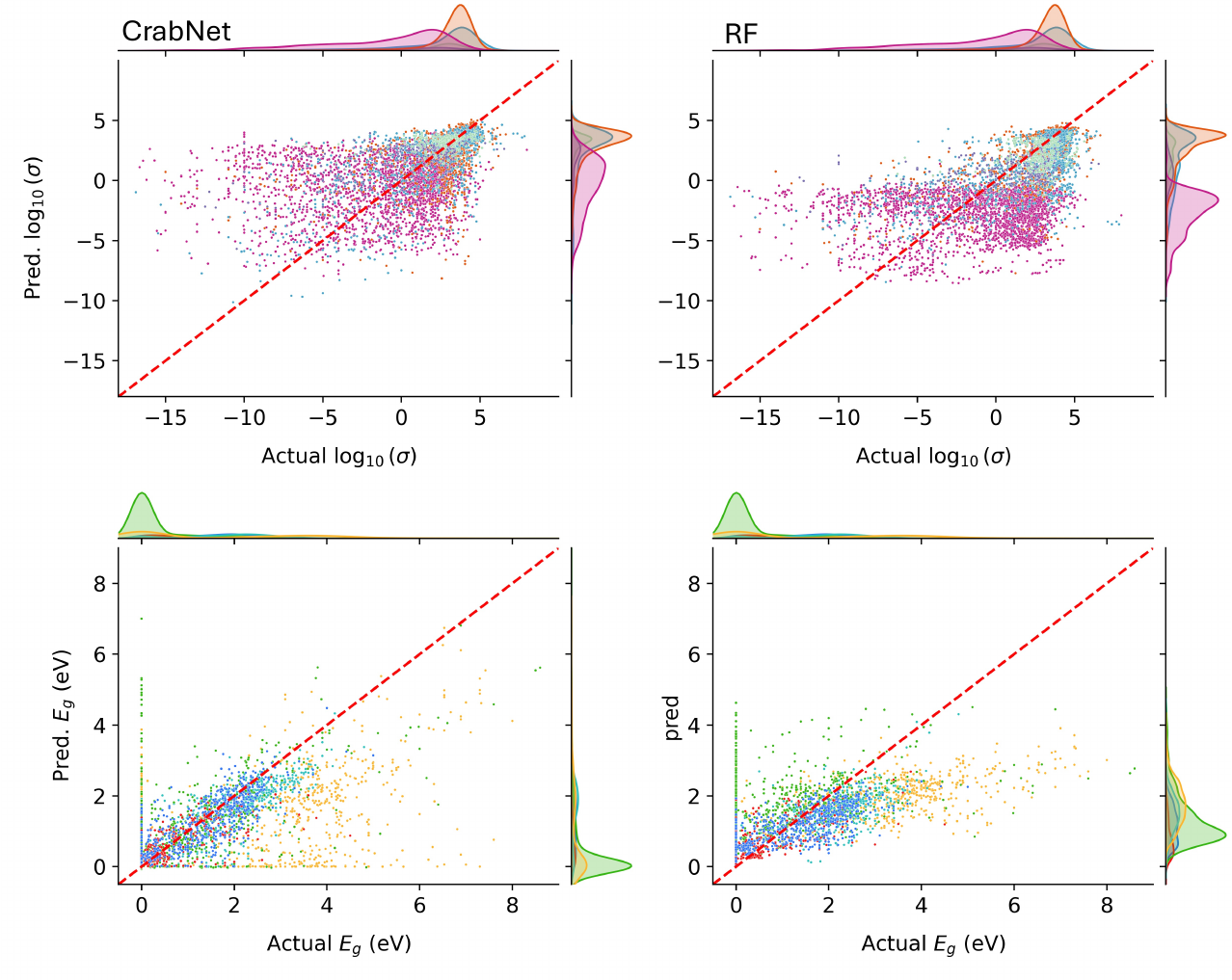}
\caption{Parity plots of ML cluster predictions under the LOCO-CV evaluation scheme.}
\label{fig:parity_loco}
\end{figure*}
where $\hat{\sigma}_{a,\theta_m}^2(\bm{x}_i)$ denotes the contribution to aleatoric uncertainty produced by the $m$-th model in the ensemble, while $\hat{\sigma}_e^2(\bm{x}_i)$ denotes the contribution to the epistemic uncertainty, obtained by computing the variance over predictions from all the models in the deep ensemble. Unlike fully Bayesian methods like Bayesian Neural Networks (BNNs)~\citep{arbel2023primerbnns}, deep ensembles approximate the Bayesian posterior by training multiple neural networks independently with different random initializations and data shuffling, providing a scalable and practical approximation to bayesian inference.
\section{LOCO-CV material clusters analysis}
\begin{figure*}[!h]
	\centering
	\includegraphics[width=0.70\textwidth]{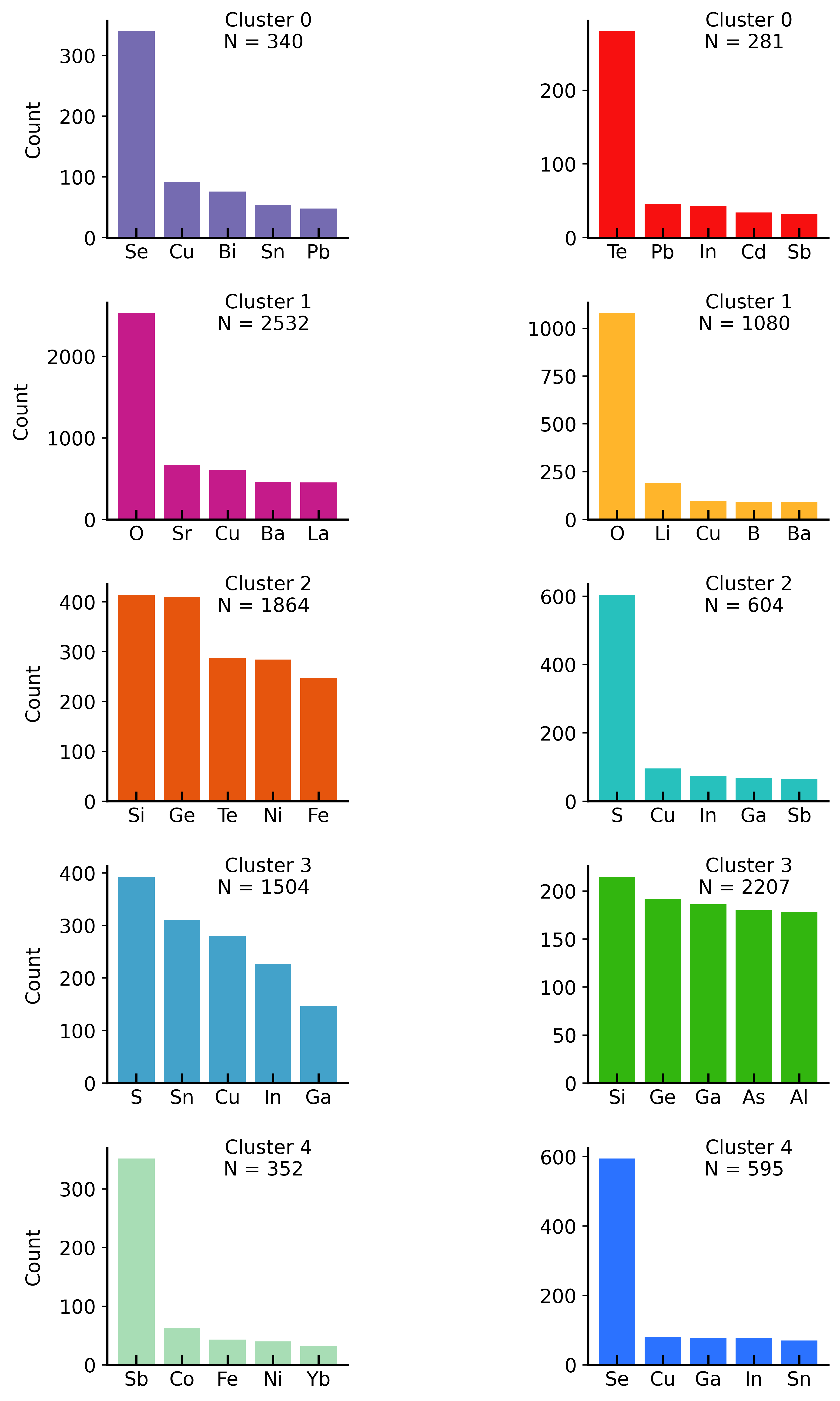}
	\caption{Top-5 element prevalence of LOCO-CV material clusters both for conductivity (\textbf{left}) and band gap (\textbf{right}) datasets.}
	\label{fig:loco_elems}
\end{figure*}

\noindent In this section, we present a more detailed analysis of material clusters generated using the LOCO-CV \citep{meredig_2018} evaluation method, as described in Sections of the main thesis. In Figure \ref{fig:parity_loco} we show parity plots related to LOCO-CV evaluation scheme, colored according to different material clusters encountered in both conductivity, and band gap datasets. In Figure \ref{fig:loco_elems} we report the top-5 element prevalence for each chemical cluster. In general, the presence of diverse, predominant elements in each cluster indicates that the clustering algorithm has successfully grouped the chemical formulas based on their composition. Moreover, the diversity of material groups suggests that the clusters effectively represent distinct regions of the chemical space, potentially capturing different types of compounds or materials.

\paragraph{Conductivity database clusters} In the case of conductivity database, \textbf{Cluster 0} (Se-Cu-Bi-Sn-Pb) \modif{consists mainly of selenium containing compounds as selenides and selenide oxide or selenide halides, while \textbf{Cluster 1} (O-Sr-Cu-Ba-La) contains oxides including sulphates and phosphates. \textbf{Cluster 2} (Si-Ge-Te-Ni-Fe) contains intermetallic compounds including borides, carbides and nitrides. \textbf{Cluster 3} (S-Sn-Cu-In-Ga) also contains intermetallic compounds along with sulphides, borides and carbides, while \textbf{Cluster 4} (Sb-Co-Fe-Ni-Yb) consists of materials all containing antimony.}

\paragraph{Band gap database clusters} For the band gap database, \textbf{Cluster 0} (Te-Pb-In-Cd-Sb) \modif{consists mainly of tellurides and lead-based compositions, while \textbf{Cluster 1} (O-Li-Cu-B-Ba) represents oxide containing compounds including sulphates and phosphates. \textbf{Cluster 2} (S-Cu-In-Ga-Sb) consists of sulphide materials, including sulphide halides. \textbf{Cluster 3} (Si-Ge-Ga-As-Al) represent intermetallics, including silicides, phosphides, carbides, borides, nitrides, while \textbf{Cluster 4} (Se-Cu-Ga-In-Sn) consists of selenide materials including selenide halides.}
\end{document}